\documentclass[aip,amsmath,amssymb,reprint]{revtex4-1}
\usepackage{graphicx}
\usepackage{dcolumn}
\usepackage{bm}
\usepackage[utf8]{inputenc}
\usepackage[T1]{fontenc}
\usepackage{mathptmx}

\usepackage{xcolor}
\usepackage{bbold}
\usepackage{float}

\usepackage[english]{fancyref}
\usepackage{amsmath}
\usepackage{amsfonts}
\usepackage{dsfont}
\usepackage{natbib}
\usepackage[a4paper, margin=2.5cm]{geometry}
\def \bfj {{\bf j}}
\def \bfi {{\bf i}}

\def \Di {{D_{\bfi}}}
\def \Dj {{D_{\bfj}}}

\def \braDi {{\langle \Di | }}

\def \ketDi {{| \Di \rangle }}

\def \ketDj {{| \Dj \rangle }}
\def \half {{\frac{1}{2}}}

\def \Eh {{\text{E}_h}}
\def \mEh {{\text{mE}_{\textrm h}}}
\newcommand{\cre}[1]{ \hat{a}_{#1}^{\dagger} }

\newcommand{\des}[1]{ \hat{a}_{#1} }

\newcommand{\rndbrk}[1]{ \left({#1}\right)}
\newcommand{\sqrbrk}[1]{ \left[{#1}\right]}
\newcommand{\crlbrk}[1]{ \left\{{#1}\right\}}

\newcommand{\excit}[2]{ \hat{E}_{#1#2} }

\newcommand{\PoseThree}[6]{\Gamma{}^{{#1}}_{{#2}}{}^{{#3}}_{{#4}}{}^{{#5}}_{{#6}}}

\newcommand{\SposeThree}[6]{\overline{\Gamma}{}^{{#1}}_{{#2}}{}^{{#3}}_{{#4}}{}^{{#5}}_{{#6}}}
\newcommand{\SposeFour}[8]{\overline{\Gamma}{}^{{#1}}_{{#2}}{}^{{#3}}_{{#4}}{}^{{#5}}_{{#6}}{}^{{#7}}_{{#8}}}

\begin{document}
\author{Robert J. Anderson}
\affiliation{Department of Physics, King's College London, Strand, London, WC2R 2LS, U.K.}
\author{Toru Shiozaki}
\affiliation{Quantum Simulation Technologies, Inc., Cambridge, MA 02139, USA}
\author{George H. Booth}
\email{george.booth@kcl.ac.uk}
\affiliation{Department of Physics, King's College London, Strand, London, WC2R 2LS, U.K.}

\title{Efficient and stochastic Multireference Perturbation Theory for Large Active Spaces within a Full Configuration Interaction Quantum Monte Carlo Framework}

\begin{abstract}
Full Configuration Interaction Quantum Monte Carlo (FCIQMC) has been effectively applied to very large configuration interaction (CI) problems, and was recently adapted for use as an active space solver and combined with orbital optimisation. In this work, we detail an approach within FCIQMC to allow for efficient sampling of fully internally-contracted multireference perturbation theories within the same stochastic framework. Schemes are described to allow for the close control over the resolution of stochastic sampling of the effective higher-body intermediates within the active space. It is found that while CASPT2 seems less amenable to a stochastic reformulation, NEVPT2 is far more stable, requiring a similar number of walkers to converge the NEVPT2 expectation values as to converge the underlying CI problem. We demonstrate the application of the stochastic approach to the computation of NEVPT2 within a (24,24) active space in a biologically relevant system, and show that small numbers of walkers are sufficient for a faithful sampling of the NEVPT2 energy to chemical accuracy, despite the active space already exceeding the limits of practicality for traditional approaches. This raises prospects of an efficient stochastic solver for multireference chemical problems requiring large active spaces, with an accurate treatment of external orbitals.
\end{abstract}
\maketitle

\section{Introduction}
Due to the inadequacies of the Hartree--Fock method and its neglect of correlated electron physics, {\em post-}Hartree-Fock methods are required in quantum chemistry to systematically move beyond this starting point towards a quantitative computational approach to the {\em ab initio} description of molecular systems.
A common partitioning of the remaining physics can be given by the loose definitions of {\em `dynamical'} and {\em `static'} contributions to the beyond-single-determinant wavefunction character.
Dynamical contributions are described as short-ranged, high-energy correlation effects arising from the instantaneous repulsion of two-electrons in close proximity forming a `Coulomb hole', present in all molecular systems.
If the correlated physics is overwhelmingly of this kind, with Hartree--Fock providing a qualitatively correct initial description, then low-order (renormalized) perturbation theories such as coupled-cluster can describe these systems with impressive accuracy. 

However, there is an increasing documented array of correlated electron systems which {\em also} exhibit so-called `static' correlation, including low-spin magnetic systems, transition states, inorganic complexes, biological active sites, and many others which exhibit strong quantum fluctuations.
This correlation emerges from the fundamental inability to describe these strong, many-body quantum fluctuations with a single determinant (as could be seen from the entanglement entropy of these systems), and therefore the qualitative breakdown of the Hartree--Fock wavefunction (single Slater determinant) as a reasonable starting point
\cite{doi:10.1063/1.5039496}.
From the single particle spectrum, these systems can often be detected from the presence of low-lying degeneracies and/or partial open-shell character, however, a definitive {\em a priori} determination of the presence (or otherwise) of static correlation is impossible in general.
In contrast to dynamic correlation, this physics requires a significant superposition of low-energy electronic configurations in the wave function description in order to account for these fluctuations, leading to the {\em multireference} approaches in quantum chemistry, generally with Complete Active Space Self-Consistent Field (CASSCF) starting points.
This relies on a complete solution to the Schr{\"o}dinger equation within a restricted, low-energy orbital subspace, which is exponentially complex with respect to the size of this subspace.

The last 15 years has seen significant growth of a number of methods which admit lower-complexity and computationally cheaper approaches to solve the static correlation of this low-energy subspace, even if the exponential scaling is only supressed, rather than eliminated.
These methods include the Density Matrix Renomalization Group 
\cite{PhysRevLett.69.2863,doi:10.1146/annurev-physchem-032210-103338}
(DMRG), selected CI methods such as Heat-bath CI (HCI)\cite{doi:10.1063/1.4998614} and 
CIPSI \cite{doi:10.1063/1.1679199,EVANGELISTI198391},
Coordinate Descent FCI \cite{Wang_2019}, 
and Full Configuration Interaction Quantum Monte Carlo (FCIQMC)\cite{doi:10.1063/1.3193710}.
The use of these methods has allowed the `active space' of this strong correlation to grow in size, to treat larger systems and with a more accurate treatment of this low-energy correlated physics.
However, while this active space treatment is required for systems with strong/static correlated physics, for quantitative accuracy this solution still needs to be the starting point for an expansion of the dynamic correlation physics, which is also present.
This has led to the development of various multireference perturbation theories, including Complete Active Space second-order Perturbation Theory (CASPT2), and $N$-Electron Valence second-order Perturbation Theory (NEVPT2) amongst others, which describe a second-order perturbative correction on top of the active space zeroth-order wave function.
Recently, DMRG techniques 
\cite{doi:10.1063/1.4976644,doi:10.1021/acs.jctc.7b00735,Wouters_2016}
have been extended to the computation of multireference perturbation theories to allow for significantly larger active spaces to be treated, while FCIQMC and selected CI methods have both been applied to orbitally-optimized CASSCF problems.
However, selected CI approaches as an active space `solver' for subsequent CASSCF or internally contracted MRPT approaches are somewhat stymied by the fact that their non-variational perturbative treatment of active space determinants are harder to consider in subsequent `external' perturbative responses.

These multireference perturbation theories are substantially more expensive than their single-reference counterparts, primarily due to their substantially larger first-order interacting space (FOIS) into which the perturbative correction is expanded, though reformulations and approximations are ongoing in attempting to reduce this cost\cite{doi:10.1063/1.4941606,VMC_PT}.
A substantial improvement in the scaling of these methods comes from the use of {\em internal contraction}.
This changes the FOIS to be the functions obtained by application of excitation operators acting directly on the (single) correlated state in the active space, rather than their action on the individual configurations in this space.
Furthermore, these excitation states can be further linearly combined to form fewer `contracted' states spanning the FOIS.
These approaches can dramatically reduce the number of free parameters required to describe the perturbative correction, and render these calculations practical with minimal loss of accuracy.

The effect of this internal contraction can be viewed as introducing effective higher-body (generally four-body) terms which need to be considered in the active space.
These higher-body effective interactions arise from the implicit virtual processes of promotion of active space electron(s) to the external space, interaction, and subsequent de-excitation back to the active space, which can be written as effective active space renormalized many-body interactions (generally up to four-body).
The perturbative resolution of these effective interactions constitute the dominant cost of these calculations, requiring the computation of higher-body reduced density matrices (RDMs) of the active space wave function.
The four-body RDM of the (full) active space wave function has a CPU cost to evaluate of $N_{\rm CASCI} N_{\rm act}^8$, where $N_{\rm act}$ denotes the number of active space orbitals, and $N_{\rm CASCI}$ denotes the number of active space configurations (and is therefore factorial in $N_{\rm act}$). This (generally) constitutes the computational bottleneck of these calculations.

In this work, we aim to describe a stochastic approach to internally-contracted multireference perturbation theory, which mitigates this cost, and therefore admits a systematically improvable approach to general strongly correlated molecules which require larger active spaces than can be currently treated.
This is approached within the framework of Full Configuration Interaction Quantum Monte Carlo.
This approach involves a dual stochastic sampling.
Firstly, the wave function in the full configuration space of active orbitals is discretized and sparsely sampled with an evolving set of `walkers' which diffuse and propagate over the course of the simulation according to a set of stochastically realized rules.
The average distribution of walkers is designed to exactly represent the exact wave function in a systematically improvable way as the number of walkers increases. 
As long as there exists sparsity in this wave function (i.e. there are a significant number of low-weighted configurations, which is generally the case), then this can be exploited to reduce the cost compared to exact techniques via this compression in terms of walkers.
As a second level of stochastic sampling, in FCIQMC the Hamiltonian operator used to determine the propagation of the walkers is also coarse-grained and stochastically sampled.
The quartic complexity of this operator is reduced to the stochastic sampling of just a single term at a time.
This mitigates the complexity of the Hamiltonian operator in the propagation, and avoids the quartic complexity of any single propagation step.

We aim to transfer the benefits of both of these stochastic compression approaches to the task of a lower-scaling, stochastic, internally-contracted multireference perturbation theory, and more specifically, the CASPT2 and strongly-contracted NEVPT2 approaches.
In this, the benefits of a sparse compressed representation of the active space wave function is maintained, while the internal contraction manifests as the requirement to sample higher-body terms between the stochastic active space walkers.
This can also be achieved via a highly sparse stochastic sampling, removing the explicit high-polynomial dependence in the computation of high-body RDMs of traditional approaches.
The RDMs and other required intermediates are accumulated in a stochastic compressed form over distributed memory, allowing for an efficient, highly-parallelised framework.
The internal contraction means that the stochastic active space dynamics has no dependence at all on the size of the external space, allowing it in principle to be applied to large systems, and distinguishing it from the uncontracted approach which requires explicit excitations into the external space and therefore has a prohibitive scaling for larger system sizes
\cite{doi:10.1063/1.4974177}.

This stochastic approach to the wave function representation, and the sparse sampling of the required higher-body terms required for the perturbative treatment allows for an efficient treatment, and we demonstrate the simulation time required to appropriately converge on accurate values, as well as the effect of increasing walker number to more finely resolve the effective perturbative interactions in the active space.
This approach can build upon the stochastic FCIQMC-CASSCF methods which have been demonstrated in previous studies, which have allowed the treatment of many tens of active space orbitals.
After benchmarking the performance and convergence characteristics of this new approach, we demonstrate the ability to solve multireference perturbation theory problems up to large active spaces of (24, 24) in this preliminary work, beyond the reach of traditional methods and sufficient to demonstrate the potential and features of the approach.

In section~\ref{sec:background}, we will recap the basics of both multireference perturbation theory and FCIQMC, where the important aspects which pertain to the subsequent stochastic perturbation theory treatment will be highlighted.
Section~\ref{sec:conditioning} will demonstrate some of the key concepts which are important to consider in the stochastic version of internally-contracted multireference perturbation theory, but are incidental in the deterministic version. This will be demonstrated via a toy system.
Section~\ref{sec:algo} details the stochastic multireference perturbation theory approach, including computational details to ensure controllable and systematically improvable sampling.
This includes the efficient datastructures to accumulate averaged intermediates, the tuneable accuracy in the stochastic resolution of higher-body terms in the active space, and the transfer of previous efficiency gains found in the semi-stochastic and replica sampling approaches to FCIQMC.
Section~\ref{sec:results} presents results, firstly compared to exact deterministic values, showing the convergence with respect to technical parameters such as the walker number, simulation iterations and sampling resolution of higher-body interactions.
Finally, we present the results of an application of stochastic NEVPT2 to the study of 24 electrons in a 24 spatial orbital active space with $\sim 7.3\times 10^{12}$ active space determinants. We conclude with a discussion of the opportunities and future extensions for this new approach.

\section{Background Theory} \label{sec:background}
In this paper, we are concerned with stochastising the evaluation of two multireference perturbation theories (MRPTs), namely CASPT2 and NEVPT2.
In both theories, the first order interacting spaces are made up of internally-contracted functions (ICFs) constructed by the application of excitation operators involving core and virtual orbital space operators to the zero-order wavefunction as a whole.
This is in constrast to the so-called uncontracted approach used in earlier work\cite{doi:10.1063/1.1679199} in which interacting spaces are made up of individual determinants external to the CAS.
Although simpler in terms of formalism, uncontracted theory suffers from unfavourable scaling, with $O\rndbrk{N_\text{CASCI}\times N_\text{core}{}^2 N_\text{virtual}{}^2}$ parameters to optimise or sample.
The uncontracted FOIS turns out to be highly overcomplete, but the more compact expansion basis of internally contracted theory is capable of an accurate characterisation of dynamic correlation with a parameter space scaling of only $O\rndbrk{N_\text{CASCI}+ N_\text{core}{}^2 N_\text{virtual}{}^2}$.

\begin{figure}[h]
\centering
\includegraphics[width=0.4\textwidth]{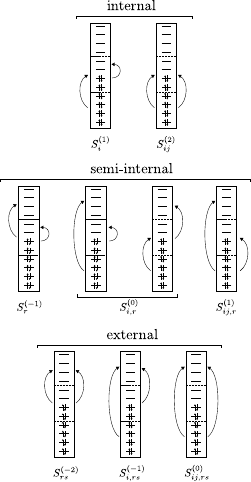}
\caption{
	Cartoon of ICFs grouped by the overall number of electrons moved to or from the active space in the first order interacting wavefunction
\label{fig:nevpt_icfs}
}
\end{figure}

It is common in the NEVPT literature to use the orbital indexing convention: $(i, j, \ldots) \in$ core space, $(a, b, \ldots) \in$ active space, and $(r, s, \ldots) \in$ virtual space, which we shall adopt.
The ICFs spanning the FOIS in CASPT and NEVPT can be orthogonally partitioned by type based on the change of electron number undergone in the active space (in superscript parentheses) and the external orbital spaces involved as illustrated in \fref{fig:nevpt_icfs}. For example, a perturber belonging to the $S_i^{\rndbrk{1}}$ subspace is constructed by the internal replacement of one electron within the CAS, and the promotion of one core electron into the CAS: $\excit{a}{i}\excit{b}{c}|\Psi^{(0)}\rangle$.
%When a single index is used to refer to a vector in the FOIS, it should be thought of as a compound index of the four orbital indices defining the perturber.

The CAS partitioning offers the benefit of a significant reduction in Hilbert space dimension, but it foregoes dynamical correlation.
MRPT2 enables the partial recovery of these neglected effects, and the internally contracted formalism facilitates this improvement without abandonning the need for larger parameter spaces.
The price of this high quality treatment of both static and dynamic correlation is the need for higher-order interactions in the effective Hamiltonian.
These interactions are expectation values of the CAS wavefunction expressed in terms of the RDMs defined in Appendix \ref{app:rdm}.
Note that the RDMs estimated in FCIQMC are in the normal-ordered form e.g. $\Gamma_{(3)} = \Gamma_{i'j'k',ijk}$, as opposed to the product-of-single-excitation form e.g. 
$\Gamma^{(3)} = \PoseThree{i'}{i}{j'}{j}{k'}{k}$ that more naturally lends itself to multireference formulations.
Regardless of operator ordering, an RDM of rank $n$ exactly determines RDMs of all ranks and operator orderings $\leq n$.

\subsection{CASPT2}
CASPT2 can be viewed as the natural extension of MP2 to a multiconfigurational zero-order wavefunction. 
The zero-order Hamiltonian is based on the generalised Fock operator
\cite{doi:10.1021/j100377a012,doi:10.1063/1.462209,FINLEY1998299}
\begin{equation}\label{casscf_fock_op}
	\hat{F} = \sum_{pq} F_{pq}\excit{p}{q}
\end{equation}
with matrix elements given by spin-averaged expectation values of a one electron operator
\begin{equation}\label{casscf_fock_matrix}
	F_{pq} \equiv \frac{1}{2}\sum_\sigma \langle\Psi^{(0)}|\{\cre{q\sigma}, [\des{p\sigma}, \hat{H}]\} |\Psi^{(0)}\rangle.
\end{equation}
The converged CASSCF state is not in general an eigenfunction of $\hat{F}$, so $\hat{F}$ is not by itself a valid choice of zero-order Hamiltonian for use in conjunction with $|\Psi^{(0)}\rangle$.
A suitable operator can be constructed by imposing that $\Psi^{(0)}$ will not interact with its orthogonal complement via the zero-order Hamiltonian.
\begin{equation}\label{caspt2_H0}
	\hat{H}^{(0)} \equiv E^{(0)} |\Psi^{(0)}\rangle\langle\Psi^{(0)}| + \hat{P}\hat{F}\hat{P}
\end{equation}
where $\hat{P}\equiv 1-|\Psi^{(0)}\rangle\langle\Psi^{(0)}|$ and $E^{(0)}$ is the expectation value of $\hat{F}$ in the CASSCF state. This ensures exact equivalence between CASPT and MPPT in the case that $\Psi^{(0)}$ is a single closed-shell determinant.

Taking the full orthogonal complement of $\Psi^{(0)}$ as the interacting space results in perturbation vectors of very high dimension, so in defining $\hat{H}_0$ it is computationally advantageous to further block-diagonalise the space orthogonal to $\Psi^{(0)}$ using projectors into: the orthogonal complement of $\Psi^{(0)}$ in the CAS ($\hat{P}_\text{K}$); the space of all single and double excitations from $\Psi^{(0)}$ which are not in the CAS ($\hat{P}_\text{SD}$); and the space of all higher-rank excitations ($\hat{P}_\text{TQ...}$).

Choosing the $N_\text{ICF}$ internally contracted functions to span a FOIS of dimension $N_\text{FOIS}$, we follow Rayleigh-Schr{\"o}dinger perturbation theory to express the CASPT1 wavefunction as the solution of a system of linear equations:
\begin{equation}\label{caspt1_wf}
	\sum_j^{N_\text{ICF}} C^{(1)}_j \langle i | \hat{H}^{(0)} - E^{(0)} | j \rangle = -\langle i | \hat{H} | \Psi^{(0)}\rangle
\end{equation}
or in matrix notation:
\begin{equation}
	\label{eq:caspt1_wf_matrix}
	\sqrbrk{\mathbf{H^{(0)}} - E^{(0)}\mathbf{S}} \vec{C}^{(1)} = -\vec{V}.
\end{equation}
where the elements of $\mathbf{H^{(0)}}$ are the matrix elements of $\hat{H}^{(0)}$ in the ICF basis and $\mathbf{S}$ is the overlap matrix of the ICFs.

%Since $N_\text{ICF}\ge N_\text{FOIS}$, \textcolor{red}{??} 
The ICFs make up an overcomplete spanning set for the FOIS, and so the overlap matrix of the ICFs is diagonalised and the eigenvectors corresponding eigenvalues smaller than a cutoff threshold, $\chi_{{\rm cut}}$, are removed, leaving $N_\text{FOIS}$ linearly independent vectors.
%$N_\text{FOIS}$ is the number of vectors remaining after these linear dependencies have been dealt with.
This diagonalisation followed by the removal of redunancies in the FOIS is represented by the $N_\text{ICF}\times N_\text{FOIS}$ matrix $\mathbf{U}$.
The singular values are denoted by
\begin{equation}
	\label{eq:caspt2_lowdin}
	\bm{\Lambda}_\mathbf{S} = \mathbf{U}^\dagger \mathbf{S} \mathbf{U}
\end{equation}
Symmetric orthogonalisation yields CASPT1 equations in an orthonormal basis
\begin{equation}\label{caspt1_wf_lowdin}
	\sqrbrk{\mathbf{H^{(0)}_S} - E^{(0)}\mathbf{1}} \vec{C_\mathbf{S}}^{(1)} = -\vec{V_\mathbf{S}}
\end{equation}
where $\mathbf{H^{(0)}_S}\equiv \rndbrk{\mathbf{U} \bm{\Lambda_\mathbf{S}}{}^\half}^\dagger \mathbf{H^{(0)}} \rndbrk{\mathbf{U} \bm{\Lambda_\mathbf{S}}{}^\half}$,

$\vec{C_\mathbf{S}}^{(1)}\equiv \rndbrk{\mathbf{U} \bm{\Lambda_\mathbf{S}}{}^\half}^\dagger \vec{C}^{(1)}$, and $\vec{V_\mathbf{S}}\equiv \rndbrk{\mathbf{U} \bm{\Lambda_\mathbf{S}}{}^\half}^\dagger\vec{V}$

Then begins the process of finding expressions for the zero-order Hamiltonian, overlap, and interaction elements in terms of the active space density matrices.
The choice of $\hat{H}^{(0)}$ ensures that inner products of this kind are non-zero only when the bra and ket belong to the same ICF type, as shown in Fig.~\ref{fig:nevpt_icfs}.
CASPT2 inner products are generally eight indexed quantities, but double occupancy of the core, and vacancy of the virtual space can be used to integrate out external space operators.
Thus, energy expressions for the $S_r^{(-1)}$ and $S_i^{(1)}$ ICF subspaces, each with three indices in the CAS, retain more nontrivial indices in the matrix elements than the other ICF types.

Taking $S_r^{(-1)}$ as an example where the ICFs take the general form $\excit{r}{c}\excit{b}{a}|\Psi^{(0)}\rangle$, the overlap and Hamiltonian elements are given by
\begin{equation}
	\label{eq:caspt2_overlap}
	S_{a'b'c',abc} = \langle \Psi^{(0)}| \excit{a'}{b'}\excit{c'}{r'}\excit{r}{c}\excit{b}{a}|\Psi^{(0)}\rangle
\end{equation}
\begin{equation}
	\label{eq:caspt2_hamiltonian}
	H^{(0)}_{a'b'c',abc} = \sum_{de}^{N_\text{act}} F_{de} \langle \Psi^{(0)}| \excit{a'}{b'}\excit{c'}{r'}\excit{d}{e}\excit{r}{c}\excit{b}{a}|\Psi^{(0)}\rangle.
\end{equation}

The commutator identity for excitation operators 
\begin{equation}
	\label{eq:rank_reduction}
	[\excit{m}{n},\excit{m'}{n'}] \equiv \excit{m}{n'}\delta_{m'n} - \excit{m'}{n}\delta_{mn'}
\end{equation}
is applied ubiquitously in the algebraic simplification of MRPT2 expectation values, and in this case it is used to exchange the positions of the virtual space indices $r$ and $r'$ such that annihilation on the CAS state can occur.

The overlap and Hamiltonian can then be expressed purely in terms of the active space density matrices
\begin{equation}
	\label{eq:caspt2_overlap_simple}
	S_{a'b'c',abc} = \SposeThree{a'}{b'}{c'}{c}{b}{a}
\end{equation}
\begin{equation}
	\label{eq:caspt2_hamiltonian_simple}
	H^{(0)}_{a'b'c',abc} = \sum_e^{N_\text{act}}\rndbrk{F_{ce} \SposeThree{a'}{b'}{c'}{e}{b}{a} + \sum_d^{N_\text{act}} F_{de}\SposeFour{a'}{b'}{c'}{c}{d}{d}{b}{a}}
\end{equation}

Simplifications arise when the Fock matrix is diagonal in the CAS, but in general this is not the case for canonical CASSCF orbitals.
We can however make use of the CASSCF wavefunction's invariance to core-core, active-active, and virtual-virtual orbital rotations, and diagonalise the three diagonal blocks of the generalised Fock matrix to work in the pseudocanonical basis.

With this rotation, Eq.~\ref{eq:caspt2_hamiltonian_simple} takes on a much simpler form:
\begin{equation}
	\label{eq:caspt2_hamiltonian_pseudocanonical}
	H^{(0)}_{a'b'c',abc} = \epsilon_c \SposeThree{a'}{b'}{c'}{c}{b}{a} + \sum_d^{N_\text{act}} \epsilon_d \SposeFour{a'}{b'}{c'}{c}{d}{d}{b}{a}
\end{equation}
where $\epsilon_p \equiv F_{pp}$.
This demonstrates that for CASPT2, the four-body active space density matrix, $\overline{\Gamma}^{(4)}$, is only required when contracted with the Fock matrix as shown above.
This is key for an efficient implementation, and we ensure that this is exploited in the FCIQMC adaptation later, by avoiding construction and storage of the full rank $\overline{\Gamma}^{(4)}$.
Instead, the \emph{CASPT2 intermediate}, which is indexed by only six active space indices is estimated, as
\begin{equation}
    \overline{\Gamma}_{(4)}F_{a'b'c',abc} \equiv \sum_d^\text{CAS} \epsilon_d \overline{\Gamma}_{a'b'c'd, abcd}. \label{eq:CASPT2_interm}
\end{equation}
Once the CAS density matrices and intermediates have been estimated, the first order interacting wavefunction can be computed, normally by employing an iterative scheme to solve what is typically a large system of equations.
The second order energy is then evaluated as the inner product of the first order wavefunction and the perturbation vector.

\subsection{Strongly contracted NEVPT2}

While CASPT2 has had many successes as a multireference perturbation theory, it does lack certain formal properties which are desirable from a method, and is susceptible to the pathology of intruder states\cite{C6SC03759C}. This led to the development of NEVPT2, which while more computationally costly than CASPT2, is strictly size consistent, and does not suffer from intruder states. This has allowed it to achieve in general higher accuracy for correlated problems of interest in the same active space.
The zero-order Hamiltonian of NEVPT2 is defined as follows:
\begin{equation}\label{eq:nevpt_h0_def}
	\hat{H}_0 = \hat{P}_{S^{(0)}}\hat{H}\hat{P}_{S^{(0)}} + \sum_{k} \hat{P}_{S^{(k)}_l} \hat{H}^D \hat{P}_{S^{(k)}_l},
\end{equation}
where $\hat{H}$ is the full, interacting, many-body Hamiltonian and the operator $\hat{P}_S$ is the projector into a vector space $S$. The CAS electron surplus of each ICF space is denoted by $k$ and the occupation pattern of each space is described by the label $l$. The CAS itself is denoted $S^{(0)}$.
The zero-order Hamiltonian definition requires that a suitable model Hamiltonian is chosen.
It has become common practice to utilise the Dyall Hamiltonian $\hat{H}^D$ for this purpose, which is bielectronic in the active space, but is purely monoelectronic in the external space.

\if 0
\begin{equation}
	\begin{split}
	\hat{H}^D \equiv \sum_i \epsilon_i \hat{a}^\dagger_{i}\hat{a}_{i}
	+ \sum_r \epsilon_r \hat{a}^\dagger_{r}\hat{a}_{r}
	\end{split}
\end{equation}
\fi

The various flavours of NEVPT2 are distinguished by which ICFs in each $S^{(k)}_l$ perturb the zero-order wavefunction $\Psi^{(0)}$.
One perturber per ICF subspace defines strongly contracted NEVPT2, many perturbers per ICF defines partially contracted NEVPT2, and the utilisation of the full space of perturbers in each ICF subspace defines uncontracted NEVPT2.

In sc-NEVPT2, the orthogonal but unnormalised perturber wavefunctions are defined by
\begin{equation}
	|\Psi_l^{(k)}\rangle = \hat{P}_{S^{(k)}_l} \hat{H} |\Psi^{(0)}\rangle \equiv |\hat{V}^{(k)}_l \Psi^{(0)}\rangle.
\end{equation}
The perturbation operators $\hat{V}^{(k)}_l$ and normalisations $N_l^{(k)}\equiv\langle\Psi_l^{(k)}|\Psi_l^{(k)}\rangle$ are expressed in full in Ref \onlinecite{doi:10.1063/1.1515317}.
Each perturber space $S_l^{(k)}$ makes a contribution 
\begin{equation}
	\label{eq:E2_slk}
	\mathcal{E}_l^{(k)} = N_l^{(k)}(E^{0}-E_l^{(k)})^{-1}
\end{equation}
to the NEVPT2 energy, where $E_l^{(k)}\equiv\langle\Psi_l^{(k)}|\hat{H}|\Psi_l^{(k)}\rangle$.
In the subsequent manipulation of the contracted subspace energy expressions, the inner products are simplified by commutation.
\begin{align}
	N_l^{(k)}E_l^{(k)} 
	=& \langle\Psi^{(0)}|\hat{V}^{(k)}_l{}^\dagger\hat{H^D}\hat{V}^{(k)}_l|\Psi^{(0)}\rangle \\
	=& \langle\Psi^{(0)}|\hat{V}^{(k)}_l{}^\dagger\hat{V}^{(k)}_l\hat{H}^D|\Psi^{(0)}\rangle \\
	&+ \langle\Psi^{(0)}|\hat{V}^{(k)}_l{}^\dagger[\hat{H}^D, \hat{V}^{(k)}_l]|\Psi^{(0)}\rangle \\
	=& N^{(k)}_l E^{(0)} + \langle\Psi^{(0)}|\hat{V}^{(k)}_l{}^\dagger[\hat{H}^D, \hat{V}^{(k)}_l]|\Psi^{(0)}\rangle
\end{align}

The expectation value of the inactive part of the Dyall Hamiltonian amounts to sums and differences of external space orbital energies to be denoted $\Delta^{(k)}_l$. 
Equation~\ref{eq:E2_slk} then becomes
\begin{align}
	\label{eq:E2_slk_delta}
	\mathcal{E}_l^{(k)} &= -N_l^{(k)}\left(
		\frac{\langle\Psi^{(0)}|\hat{V}^{(k)}_l{}^\dagger[\hat{H}^v, \hat{V}^{(k)}_l]|\Psi^{(0)}\rangle}{N^{(k)}_l} + \Delta^{(k)}_l
	\right)^{-1}  \\
	&\equiv
	-N_l^{(k)}\rndbrk{\frac{h^{(k)}_l}{N^{(k)}_l} + \Delta^{(k)}_l}^{-1}
\end{align}

The energy expressions for the $S_r^{(-1)}$ and $S_i^{(1)}$ contracted subspaces will once again depend on the active space RDMs of highest rank, since they contain the most active space excitation operators: three from the product of $\hat{V}$ and $\hat{V}^\dagger$ and two from the interactions of the active space Hamiltonian. This apparently rank five density matrix expression in fact only depends on the rank four density matrix after rank reduction identity of Eq.~\ref{eq:rank_reduction} is effected.

Furthermore, $\overline{\Gamma}^{(4)}$ only occurs in two transpositionally nonequivalent contractions in the $E_r^{(-1)}$ and $E_i^{(1)}$ expressions, specifically:
\begin{align}
	\label{eq:nevpt2_intermediate_def_1}
	\overline{G^A}_{a' b' c', a b c} &= \sum_{def}^{N_\text{act}} \langle de | fa \rangle \SposeFour{c'}{a'}{b'}{b}{d}{f}{e}{c}; \\
	\label{eq:nevpt2_intermediate_def_2}
	\overline{G^B}_{a' b' c', a b c} &= \sum_{def}^{N_\text{act}} \langle dc | fe \rangle \SposeFour{c'}{a'}{b'}{b}{d}{f}{a}{e}.
\end{align}
These contractions, which will hereafter be referred to as the \emph{NEVPT2 intermediates}, and are critical to the efficiency in the same fashion as the CASPT2 intermediate, conveniently reducing the memory scaling of attempts to estimate the NEVPT2 energy corrections which depend on $\overline{\Gamma}^{(4)}$.
However, no approximation to $\overline{\Gamma}^{(4)}$ is made, resulting in no further approximation to the NEVPT2 energy beyond the FCIQMC sampling.
However, the contraction and estimation of these quantities constitutes the computational bottleneck of the full algorithm.
Rank-reducing approximations to aid efficiency have been developed for deterministic ic-MRPT, with varying degrees of success due to the reintroduction of intruder-state problems, but may warrant consideration in future work.
\cite{doi:10.1063/1.3132922, doi:10.1021/acs.jctc.6b00714}

\subsection{FCIQMC in brief}
Full configuration interaction quantum monte carlo\cite{doi:10.1063/1.3193710} solves the correlation problem by representing the many-body wavefunction in Slater determinant space as a population of discrete, signed \emph{walkers}, which provide a coarse-grained snapshot of the FCI amplitudes. The exact FCI wavefunction can be represented as
\begin{equation}
|\Psi \rangle = \sum_{\bf i} C_{\bf i} |D_{\bf i} \rangle ,
\end{equation}
where ${\bf i}$ labels a many-body configuration in the chosen orbital basis.
Within FCIQMC, the exact CI coefficients are replaced by a discretized, sparsely sampled walker weights, i.e. $C_{\bf i} \rightarrow {\tilde{C}_{\bf i}}(\tau)$, which are overwhelmingly zero at any single iteration.
%The walker population iteratively evolves under the stochastic application of the shifted propagator $1-\tau(\hat{H}-S_0)$.
An instantaneous snapshot of the fluctuating distribution of walkers therefore generally provides a rather poor discription of the wavefunction, but with a sufficent number of walkers in the system, the long time average walker distribution can provide accurate expectation values, often at a small fraction of the memory and CPU cost of deterministic analogues.
With a suitably small timestep $\Delta\tau$ and given an initial guess vector of walkers representing $|\Psi(\tau=0)\rangle$, the FCIQMC walker population evolves under the stochastic application of the propagator
\begin{equation}
	\left(1-\Delta\tau\hat{H}\right)^N |\Psi(\tau=0)\rangle \propto |\Psi_0 \rangle ,
\end{equation}
which will tend after normalisation toward the exact ground state $|\Psi_0\rangle$ of a many-body Hamiltonian $\hat{H}$ in the limit of large $N$, provided that $\langle\Psi_0|\Psi(\tau=0)\rangle \neq 0$.
Shifting the Hamiltonian by an estimate $S_0$ of the ground state eigenvalue yields a norm-conserving propagator, which then does not result in a net growth or decay in the walker number.% by maximising the exponential reduction of the overlap with excited states at each iteration.

This can be done within a determinant space with restricted (e.g. CASCI) orbital space occupations.
Such sparse samplings of the active space wavefunction allows for the estimation of $\Gamma_{(2)}$.
This has been implemented previously and deployed in stochastic orbital optimisation with CASSCF~\cite{doi:10.1021/acs.jctc.5b00917}, which has been applied to active orbital spaces which correlate 32 electrons in 29 spatial orbitals~\cite{doi:10.1021/acs.jctc.5b01190}.

The unaltered FCIQMC algorithm often requires a large number of walkers in order to overcome the fermion sign problem~\cite{Spencer_2012}, and so it is a practical necessity to employ the \emph{initiator} approximation~\cite{doi:10.1063/1.3302277}.
This introduces a small and systematically improvable error in the sampling by effectively setting to zero the Hamiltonian matrix elements connecting low weight determinants to the space of unoccupied determinants.
The initiator approximation brings the advantage that the overall number of walkers required to reach high accuracy is dramatically decreased.
In this way, the number of walkers sampling the wavefunction is now a parameter which can be increased to converge all the many-body effects of the system.
The recent innovation of correcting for the introduction of systematic error from the initiator approximation via PT2 has shown impressive results~\cite{doi:10.1063/1.5037923}, however this is only generally applicable to the energy estimator and more difficult to use as a correction to properties or estimations $\Gamma_{(n)}$.

More details on the FCIQMC algorithm can be found in 
Refs~\onlinecite{doi:10.1063/1.3193710,
doi:10.1063/1.4904313,
booth/nature,
Blunt/2015,
doi:10.1063/1.3302277,
doi:10.1063/1.4986963,
doi:10.1063/1.4920975}.
All FCIQMC development in this work was undertaken in the NECI package\cite{neci}. Modifications were made to PySCF\cite{pyscf} for the stochastic NEVPT2, while an interface to the BAGEL program\cite{doi:10.1002/wcms.1331} was developed in order to realise stochastic CASSCF and CASPT2.

\section{Numerical conditioning and noise}
\label{sec:conditioning}
Expectation values estimated by FCIQMC and other stochastic quantum chemistry approaches necessarily contain random error.
Therefore, before setting out to stochastise any theory, it is prudent to investigate the effect of this random error in a controlled setting, in terms of its effect on the stability of the algorithm, the random errors in the final quantities, and the potential systematic biases that the random errors introduce.
The original FCIQMC algorithm only involves linear operations on random variables, and therefore is essentially free of systematic bias in the sampled quantities (CI coefficients as well as projected energy estimates), up to the effect of the initiator approximation.
The MRPT algorithms involve accumulating and sampling higher-rank quantities. The act of contracting these quantities down to get scalar expectation values can cancel much of the random error in the individually sampled values, resulting in random error in the final quantities smaller than those of the individual elements.

However, while the individual many-body density matrices can be sampled in an unbiased fashion without systematic error in FCIQMC (see subsection \ref{rdm_fciqmc}), the MRPT algorithms also involve non-linear operations on these stochastic variables. This introduces covariances between the stochastic quantities, and can manifest as a systematic error in the final quantities which depends on the magnitude of the random error in the original variables. In some algorithms, this non-linear bias can be the dominant source of error, while in others, such as the FCIQMC-based CASSCF implementations or excited state FCIQMC algorithm, it can be relatively well controlled and not cause significant error. However, the source of these non-linear operations in MRPT theories and their potential biases will need to be carefully considered.

Specifically, matrix inversions are problematic when the matrix elements are estimated by a stochastic process, and since perturbation theories always feature a denominator of energy differences (e.g. Eqs.~\ref{eq:caspt1_wf_matrix} and \ref{eq:E2_slk}), their compatibility with Monte Carlo methods must be the subject of close scrutiny. 
A particularly noteworthy difference between the two MRPTs under consideration in this work is that the internally contracted functions of sc-NEVPT2 are orthogonal by construction, while those of CASPT2 are not, and must be L{\"o}wdin-orthogonalised by inversion of the overlap matrix before the ICF energies can be evaluated.
The full space of ICFs contains linear dependencies, and so the inversion step is preceded by an eigendecomposition, which facilitates the elimination of ICFs corresponding to small eigenvalues, before they can cause singularities (or numerically problematic near-singularities) in the orthogonalisation step. This is a highly non-linear operation, and if the quality of the results is sensitive to the accuracy of these small eigenvalues, then numerical instabilities as well as biases can result.
Diagonalizing stochastically derived, non-orthogonal subspace Hamiltonians has already been investigated more generally in the context of FCIQMC, for the computation of excited state, thermal and spectral properties, where the numerical conditioning was found to require careful consideration
\cite{PhysRevLett.115.050603, PhysRevB.98.085118}.

\begin{figure}[h]
	\centering
	\includegraphics[width=0.5\textwidth]{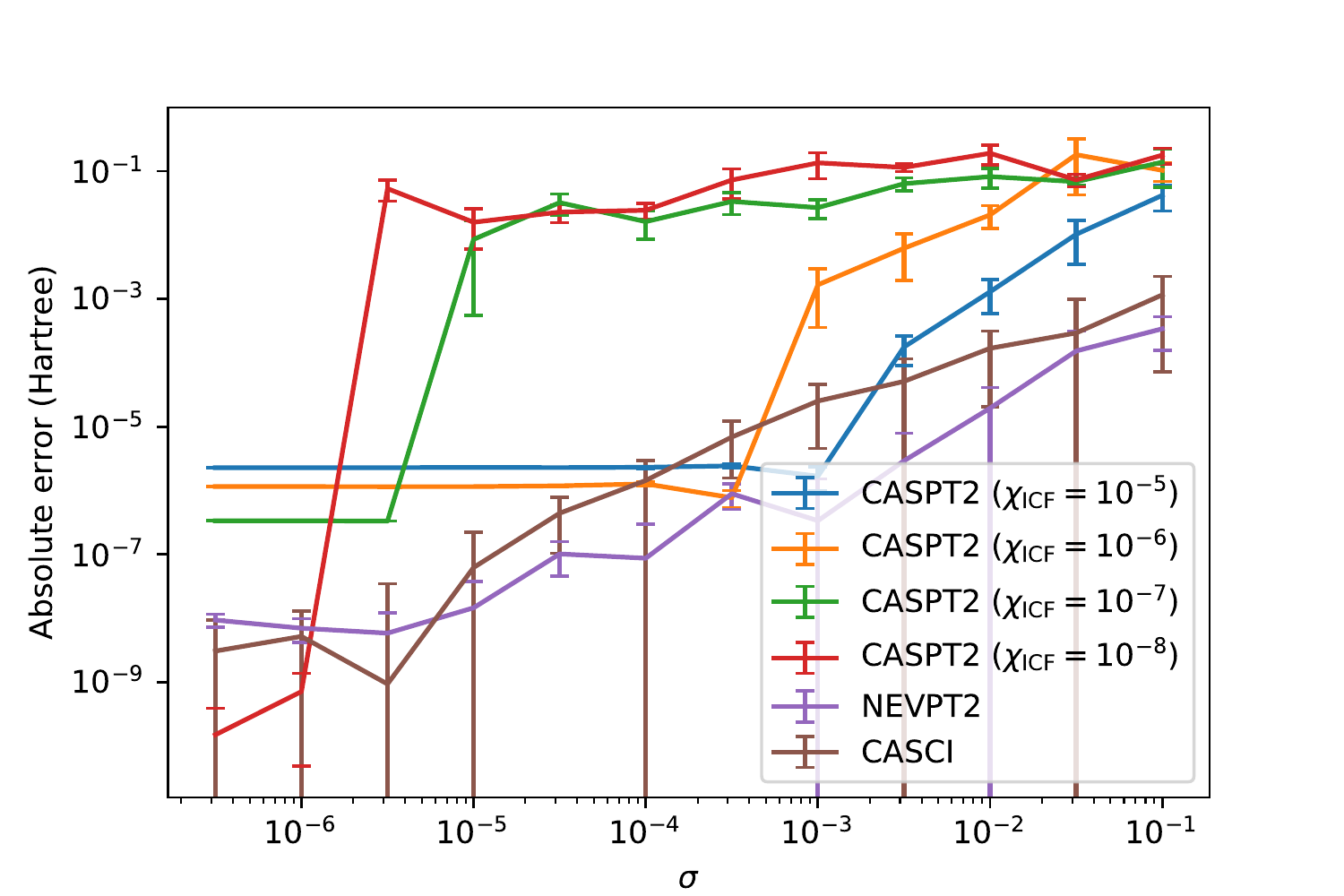}
	\caption{The noisy wavefunction experiment for a simple system, namely the $\pi$ space of N$_2$ in a cc-pVDZ basis set. Shown is the absolute energy error relative to deterministic CASPT2 (with a 10$^{-9}$ ICF overlap threshold) as a function of the magnitude of the Gaussian aberration added to the exact ground state eigenfunction of the CAS Hamiltonian.
	The overlap matrix eigenvalue cutoffs for noisy CASPT2 are parenthesised in the legend.
    CASCI and NEVPT2 energy errors are also included but are not subject to thresholds, and represent the errors due to the same relative Gaussian noise.
	\label{fig:noise_experiments}}
\end{figure}

In order to provide a simple and controllable test of the effect of noisy CI amplitudes on the performance and magnitude of the biases of the MR perturbation theories considered in this work, we computed the exact CASSCF wavefunction for a small system, then added noise to it based on a normal distribution with a width proportional to the magnitude of the exact CI vector elements,
\begin{equation}
	\widetilde{C}_\bfi = \mathcal{N}(C_\bfi,\sigma |C_\bfi|)
\end{equation}
The RDMs and intermediates were then deterministically constructed from this noisy wavefunction before the MRPT2 energies could be evaluated.
This procedure was repeated multiple times from which the mean and standard error could be plotted.
Figure \ref{fig:noise_experiments} shows the energy error of CASPT2 calculations as a function of wavefunction noise for a selection of ICF overlap cutoff parameters.
It is clear from the results of this experiment that the smaller the $\chi_\text{ICF}$ cutoff value, the smaller the noise in the wavefunction must be to evade catastrophic conditioning related errors in CASPT2.
However, it seems that CASPT2 energy estimates based on noisier CAS wavefunctions can be made numerically well conditioned with only a modest sacrifice in best case accuracy due to the requirement of larger $\chi_\text{ICF}$ thresholds introducing systematic errors.

NEVPT2 on the other hand is very resilient to the artificially introduced noise, with results comparable in accuracy and precision to the CASSCF energy, which has already been found very amenable to a robust stochastic formulation
\cite{doi:10.1021/acs.jctc.5b00917,doi:10.1021/acs.jctc.5b01190}.
This is largely due to the manifest orthogonality of the FOIS construction.
This bodes well for the stochastisation of NEVPT2 with FCIQMC, although it should be pointed out that this experiment does not model the sparse estimation of RDM elements.
Also, the RDM estimates are biased, because different realisations (replicas) of the bra and ket vectors were not used, as would be the case in an actual FCIQMC calculation.
However, this toy example serves as a useful experiment which illustrates some of the factors that will need to be considered in the stochastic algorithms, and where we expect NEVPT2 to exhibit a less severe bias with respect to random error than CASPT2.

\section{Stochastic ic-MRPT algorithm} \label{sec:algo}
From the point of view of FCIQMC implementation, the adoption of an internally contracted formalism has the advantage that no modification of the master equation that governs the walker dynamics is required.
There are no additional source terms, and no need to store walkers on determinants outside of the active space.
The stochastic ic-MRPT algorithm instead calls for the estimation of higher-body quantities in the active space, and thus a generalisation of the two-body RDM subalgorithm to enable the direct sampling of the three- and four-body density matrices is required.
Spectating the dynamics of the stochastic wavefunction in order to sample these quantities comes with its own challenges, which are detailed in this section.

\subsection{Density matrix accumulation in FCIQMC}
\label{rdm_fciqmc}
The two-body RDM of a wavefunction with many-body amplitudes given by $C_\bfi$ is defined as
\begin{equation}
	\Gamma_{ij,kl} = \sum_{\bfi\bfj} C_\bfi^*C_\bfj \braDi \cre{i}\cre{j}\des{l}\des{k} \ketDj ,
\end{equation}
and is critical for computing a number of molecular electronic properties: most importantly, the energy.
It also finds application in determining the rotations between active and external orbital subspaces in a CASSCF optimisation.

FCIQMC has been adapted to estimate these quantities \cite{doi:10.1063/1.4986963,doi:10.1063/1.4904313}.
Accumulations of the \emph{diagonal} $\Gamma_{ij,ij}$ elements of this object correspond to sums over the instantaneous magnitudes of individual determinants.
Instances of these accumulations are called ``diagonal contributions''.
Each occupied determinant relates to ${N_\text{elec} \choose 2}$ diagonal elements of $\Gamma_{(2)}$.
Naively, this would entail an $O(N_\text{elec}{}^2)$ loop every iteration, however the block averaging strategy detailed in the references~\cite{doi:10.1063/1.4904313,doi:10.1063/1.4986963}, ensures that such a loop need only be undertaken when the determinant becomes unoccupied, or the end of an RDM averaging block of Monte Carlo cycles is reached.
Since the RDM is quadratic in the wavefunction, a systematic bias is necessarily introduced as a consequence of the stochasticity~\cite{doi:10.1063/1.4904313}.
This bias affects the diagonal elements of the RDM particularly strongly, and since these elements also implicitly determine the off-diagonal elements through normalisation by the trace, every RDM element will suffer from this bias if measures are not taken to avert it.

The solution is found in the introduction of another statistically independent population of walkers and have the two \emph{replicas} evolving concurrently.
As a result, the stochastic RDM sampling is \emph{unbiased} by taking one CI coefficient from each replica
\begin{equation}
	\label{eq:rdm_contrib}
	\Delta \Gamma_{ij,kl} = \sum_{\bfi\bfj} C_\bfi^{(\text{replica 1})}{}^*C_\bfj^{(\text{replica 2})} \braDi \cre{i}\cre{j}\des{l}\des{k} \ketDj
\end{equation}
The second-quantised matrix element in Eq.~\ref{eq:rdm_contrib} amounts to a parity factor of $\pm1$ reflecting whether an even or odd number of antisymmetric exchanges of operators is required in order to eliminate it.

For the \emph{off-diagonal} elements of $\Gamma_{(2)}$, we must consider pairs of occupied determinants.
Rather than explicitly iterating over connected pairs of occupied determinants at every Monte Carlo cycle to accumulate the RDM elements, in general a hybrid approach is employed which incorporates the stochastically sampled connections from the spawning step of the algorithm for the low-occupation determinants.
During the spawning stage of the FCIQMC algorithm, every occupied determinant generates some number of other determinants to which it may be connected through an operator of interest.
In the case of pure FCIQMC dynamics, the operator in question is the Hamiltonian, and the connections generated for the purposes of evolving the walker population according to the propagator also make contributions to $\Gamma^{(2)}$.
The number of attempts is equal to the walker occupation on the parent determinant on average, with each realisation using a stochastic round to determine an integral number of attempts.
This procedure admits a finer sampling of RDM elements involving highly occupied determinants.

In the case of purely stochastic propagation, all off-diagonal elements are sampled this way, apart from single and double excitations of the reference determinant.
These RDM contributions are always taken into account with block averaging and a deterministic loop, since they are expected to have high walker occupations, and in the case of HF orbitals, the single excitations would never be sampled by the excitation generator due to Brillouin's theorem.
There is no additional stochastisation involved in the block averaged contributions to RDM elements, and their errors are due only to the random and systematic (initiator) errors in the CI coeffcients caused by discretisation.
This also holds true for errors in the diagonal elements of the RDM, which only receive contributions from block averaging.

When a deterministic subspace is defined to aid efficiency \cite{doi:10.1063/1.4920975},
the connections between its determinants are enumerated at the initialisation of the semi-stochastic propagation, and the resulting sparse map is used to efficiently perform the loop over connections between these generally highly-occupied determinants, and to include their RDM contributions.

The RDMs can be divided into elements where all indices are unique, and ones where at least one index is repeated.
The contributions to these two classes are algorithmically separated into {\emph direct} contributions for the former class, and {\emph promotion} operations for the latter.
Direct contributions come from the product of amplitudes on determinants separated by an excitation level equal to the rank of the RDM being accumulated, and for the sampling therefore require the random excitation of determinants of this rank.
Promotion contributions on the other hand derive from lower-body excitations, and involve considering all common indices in these low-body excitations.

For instance, double excitations such as $\ketDi \rightarrow \cre{i}\cre{j}\des{l}\des{k}\ketDi\equiv\ketDj$ call for a contribution to just one off-diagonal $\Gamma_{(2)}$ element, $\Delta\Gamma_{ij,kl}$ where $i\neq k$ and $j\neq l$, requiring a direct sampling of this element.
If the connection was generated stochastically, the contribution is
\begin{equation}
	\label{eq:rdm2_update}
	C_\bfi^{\text{(replica 1)}}{}^*C_\bfj^{\text{(replica 2)}}/p_{\text{gen }\geq 1}\rndbrk{\bfj\leftarrow\bfi}
\end{equation}
where $p_{\text{gen }\geq 1}\rndbrk{\bfj\leftarrow\bfi}$ is the probability that at least one spawned walker was generated to sample the connection.
The calculation of this probability is necessary to allow for the consolidation of all spawned contributions bound for the same destination determinant into a single $\Delta C_\bfj$, but while still keeping track of the source determinants responsible for each individual contribution.
This ensures that $\Gamma_{(2)}$ can be correctly computed using the instantaneous walker occupations of the destination determinants, before all spawned weights take effect, without the need to ever retain a copy of the walker list from the previous MC cycle.

However, single excitations such as $\ketDi \rightarrow \cre{i}\des{j}\ketDi\equiv\ketDj$ where $i\neq j$ however, demand contributions to multiple elements of $\Gamma_{(2)}$ via \emph{promotion}, namely $\Delta\Gamma_{ik,jk}$ for all $N_\text{elec}-1$ orbitals $k$ occupied in both $\ketDi$ and $\ketDj$.
This promotion of excitations of lower rank than that of the sampled RDM is an insignificant additional expense in the estimation of $\Gamma_{(2)}$, but in the case of higher-body RDMs, this becomes an important consideration.

Note that with efficient stochastic excitation generation, one- and two-body connections corresponding to vanishing Hamiltonian matrix elements never result in a spawning event, and this can mean the introduction of an error in the RDM sampling.
Previous investigations~\cite{doi:10.1063/1.4904313} however have found this error to be negligibly small.
FCIQMC RDM elements are stored and communicated much like FCIQMC walkers, in that a sparse, distributed array is implemented by hash tables, and the responsibility for storing each element is delegated among the communicating processes.
RDM elements are subscripted by pairs of ascending order spin orbital index arrays, making use of permutational symmetry.
Spin averaging, and the symmetry of hermiticity is not acted upon until the end of the calculation, where elements are averaged over each side of the diagonal.

FCIQMC RDMs $\widetilde{\Gamma}_{(n)}$ are therefore systematically improvable stochastic estimates of the exact RDMs $\Gamma_{(n)}$. An FCIQMC RDM estimate tends toward the exact RDM in the limit of many sampling iterations and large walker number.
\begin{equation}
	\label{eq:stoch_rdm_limits}
	\lim_{N_\text{W}\rightarrow\infty} \lim_{N_\text{iter}\rightarrow\infty} \widetilde{\Gamma}_{(n)}(N_\text{W}, N_\text{iter}) = \Gamma_{(n)}
\end{equation}

For the purpose of improving the walker distribution toward a faithful stochastic realisation of the exact wavefunction, these two limits have distinct functions:
\begin{itemize}
	\item $N_\text{W}$ is increased to resolve the systematic error associated by the initiator approximation, since the exact many-body Hamiltonian is recovered in that limit
	\item $N_\text{iter}$ is increased (after the initial equilibration phase) in order to fulfil the requirement that the exact wavefunction is only recovered as the \emph{average} over many snapshots of the walker distribution.
\end{itemize}

However when employing such a stochastic wavefunction in the estimation of quantum mechanical expectation values, these parameters (along with others including the size of the deterministic space $N_\text{SS}$, and the number of spawns attempted per walker) begin to play overlapping roles with regard to systematic improvability.
Therefore, when $N_\text{W}$ and $N_\text{iter}$ are large enough to resolve the wavefunction to a high degree of accuracy, a question arises over the optimal means by which to improve the RDMs toward the highest attainable accuracy for the given sampled wavefunction.
This is an important consideration for the present study, in which the semi-stochastic adaptation is found to be an efficient alternative to increasing $N_W$ or $N_\text{iter}$.

Endowing the FCIQMC algorithm with the ability to compute the higher body reduced density matrices required for MRPT2 calculations requires some novel adaptations which are presented here.

\subsection{Higher-body excitation generation}
In order to sample expectation values within the FCIQMC dynamic which depend on $\Gamma_{(3)}$, $\Gamma_{(4)}$, and any required contracted arrays which may be computed therefrom, it is necessary to stochastically generate connections of rank three and four.
This requires a modification of the approach used for the two-body density matrix, in which the one- and two-body excitations necessary for the stochastic propagation of the wavefunction is sufficient.

One aspect of this modification is the explicit generation of three- and four-body excitations, which faciltate the sampling of $C_\bfi^*C_\bfj$ products that would never be selected by the propagating excitation generators.
We call the connections resulting from these processes \emph{ghost} spawnings, since they serve only to convey the instantaneous walker occupation of the source determinant such that the corresponding RDM contributions can be made, but ultimately have no effect on the walker occupation of the destination determinant.

It is important to note that the CASPT2 intermediate in the pseudocanonical basis of Eq.~\ref{eq:caspt2_hamiltonian_pseudocanonical} is a contraction over the $\Gamma_{(4)}$ elements which have at least one common orbital among the creation and annihilation operators, and so in this instance, only the three-body excitations are required to sample the intermediate. % exactly (on average, and in the large walker number limit).

Higher-body excitation generation is implemented by first stochastically rounding the number of walkers on a given determinant about some predefined value---denoted $N_W/N_\text{ghost}$---to determine the average number of ghost walker generation attempts per walker. $N_W/N_\text{ghost}$ therefore determines the probability of a unit-weight walker attempting a three- or four-body excitation to find a contribution to parts of $\Gamma_{(3)}$ or $\Gamma_{(4)}$ which are not given by the promotion of propagating spawns.
This parameter is introduced to allow for the coarseness of the three- and four- body excitation sampling to be controlled (which is greatly more expensive than normal excitation generation) and allow for the investigation of the importance of a fine sampling of the (primarily) unique index higher-body RDM contributions.

The generation of 3- or 4-body excitations from occupied determinants is carried out by a two step algorithm.
First an unconstrained, unweighted single/double excitation is generated in two separate selections.
The (1 or 2) electrons selected from the source determinant are effectively removed so they will not be subsequently selected.
Similarly, the (1 or 2) unoccupied orbitals selected from the virtual complement of the source determinant are also effectively removed.
The \emph{depleted} $N-(1 \text{ or } 2)$ electron determinant and $M-N-(1 \text{ or } 2)$ orbital virtual complement form the basis for the two remaining selections.
The second selection of occupied orbitals is constrained to result in an overall spin that can possibly be negated by the final selection.
e.g. if the initial selection has not conserved spin symmetry, choosing an $\alpha$ electron and a $\beta$ virtual orbital, the possibility of drawing an $\alpha$ pair from the depleted determinant must be zero, otherwise there would be no possible way to conserve symmetry in the final selection of virtual orbitals.
The second selection of unoccupied orbitals is constrained by both spin and point group symmetry.
A conserving pair in the virtual complement is randomly drawn, and the overall $(3 \text{ or } 4)$ body excitation is obtained by sorting the selected orbital indices.

The normalized generation probability, $p_\text{gen}$, must also be computed by the excitation generator.
The two-step nature of the algorithm causes complications, since the drawn excitation could have occurred in multiple ways, each with differing probabilities.
e.g. if a three-body excitation is drawn as a single excitation $\crlbrk{a}\leftarrow \crlbrk{i}$, followed by a constrained double excitation $\crlbrk{b,c}\leftarrow \crlbrk{j, k}$, then the overall probability of that particular \emph{selection route} is $p_\text{gen}\rndbrk{\crlbrk{b,c}\leftarrow \crlbrk{j, k} | \crlbrk{a}\leftarrow \crlbrk{i}}$, the conditional probability of selecting $\crlbrk{b,c}\leftarrow \crlbrk{j, k}$, given the previous selection of $\crlbrk{a}\leftarrow \crlbrk{i}$.

In this example, the true $p_\text{gen}\rndbrk{\crlbrk{a, b, c}\leftarrow \crlbrk{i, j, k}}$ is the sum over all selection route probabilities.
\begin{align}
	\label{eq:p_gen_three}
	\begin{split}
		p_\text{gen}&\rndbrk{\crlbrk{a, b, c}\leftarrow \crlbrk{i, j, k}} = \\
		&p_\text{gen}\rndbrk{\crlbrk{b,c}\leftarrow \crlbrk{j, k} | \crlbrk{a}\leftarrow \crlbrk{i}}\\+
		&p_\text{gen}\rndbrk{\crlbrk{b,c}\leftarrow \crlbrk{i, j} | \crlbrk{a}\leftarrow \crlbrk{k}}\\+
		&p_\text{gen}\rndbrk{\crlbrk{b,c}\leftarrow \crlbrk{k, i} | \crlbrk{a}\leftarrow \crlbrk{j}}\\+
		&p_\text{gen}\rndbrk{\crlbrk{a,b}\leftarrow \crlbrk{j, k} | \crlbrk{c}\leftarrow \crlbrk{i}}\\+
		&p_\text{gen}\rndbrk{\crlbrk{a,b}\leftarrow \crlbrk{i, j} | \crlbrk{c}\leftarrow \crlbrk{k}}\\+
		&p_\text{gen}\rndbrk{\crlbrk{a,b}\leftarrow \crlbrk{k, i} | \crlbrk{c}\leftarrow \crlbrk{j}}\\+
		&p_\text{gen}\rndbrk{\crlbrk{a,c}\leftarrow \crlbrk{j, k} | \crlbrk{b}\leftarrow \crlbrk{i}}\\+
		&p_\text{gen}\rndbrk{\crlbrk{a,c}\leftarrow \crlbrk{i, j} | \crlbrk{b}\leftarrow \crlbrk{k}}\\+
		&p_\text{gen}\rndbrk{\crlbrk{a,c}\leftarrow \crlbrk{k, i} | \crlbrk{b}\leftarrow \crlbrk{j}}
	\end{split}
\end{align}
In the corresponding sum for the four-body case, there are ${4\choose 2}^2=36$ selection route probabilities that must be retrospectively evaluated for a successfully generated excitation.

Since the ghost walkers by definition have no effect on the stored walker occupations, there is no need for the consolidation procedure leading to Eq.~\eqref{eq:rdm2_update}, and the three- and four-body ghost walkers are processed sequentially even if they correspond to the same destination determinant.
Therefore for ghost spawning, the magnitude of the RDM contribution is simply given by
\begin{equation}
	\label{eq:rdm2_update}
	C_\bfi^{\text{(replica 1)}}{}^*C_\bfj^{\text{(replica 2)}}/p_\text{gen}\rndbrk{\bfj\leftarrow\bfi}.
\end{equation}

In the NEVPT2 case where 3- and 4-body excitations make contributions to the required contracted intermediates of Eqs. \ref{eq:nevpt2_intermediate_def_1} and \ref{eq:nevpt2_intermediate_def_2},
the choice of this excitation level is randomly chosen before the orbital selection algorithm, with the ratio of probabilities being optimized initially to reflect the number of determinants of each excitation level accessible from the reference determinant in the target spin sector.
Within a selected excitation level case, the generation is constrained only by symmetry considerations, and does not attempt importance sampling by giving precedence to excitations with the most effect on the RDMs or final MRPT2 energy, as has been found efficient in recent developments in two-body sampling in FCIQMC
\cite{doi:10.1063/1.4986963,doi:10.1063/1.4904313}.
The use of Fock matrix elements and electron repulsion integrals to inform an optimal importance sampling of MRPT2 higher-body excitation generation is an avenue for further research which has the prospect of substantially improving efficiency.

\subsection{Diagonal elements and promotion}
The contracted $\Gamma_{(4)}$ sampling introduces a bottleneck whereby the cost of promotion (introduced in subsection~\ref{rdm_fciqmc}) increases to scale as
$\mathcal{O}\sqrbrk{N_\text{elec}{}^3}$ for single excitations,
$\mathcal{O}\sqrbrk{N_\text{elec}{}^2}$ for double excitations, and as
$\mathcal{O}\sqrbrk{N_\text{elec}{}}$ for triple excitations.
These promotions require loops over the combinations of occupied orbitals held in common by the bra and the ket determinants in the contributing excitation.
Each unique combination of these occupied orbitals, which constitute repeated index elements in $\Gamma_{(4)}$ are---like the four-body excitations---used for on-the-fly contraction to the CASPT2 or NEVPT2 intermediates instead of accumulation within $\Gamma_{(4)}$ directly.
This results in additional contributions to the six-indexed intermediates which must be communicated to the process on which they are sparsely accumulated.

Diagonal elements are actually the most extreme instance of promotion, since $\Gamma_{(4)}$ requires a loop over all 4-tuples of occupied orbitals.
This cost is mitigated by the infrequent accumulation of diagonal contributions thanks to block averaging of walker occupations.
However, this quartic cost is still incurred when a sampling block ends or whenever a walker dies, creating the need for very large communication buffers which can lead to the exhaustion of memory space, and also leads to a large communication overhead. 
A solution is to instead accumulate the diagonal elements of the higher-body RDMs in a dense format on each process privately, only to perform the communication step once at the end of sampling, as this only entails $\mathcal{O}\sqrbrk{n_\text{orb}{}^4}$ memory overhead.
In this approach, accumulation of diagonal elements is done without communication, and all spin averaging and intermediate contributions pertaining to the diagonal RDM elements can be left till the end of the calculation.
This measure was effective in reducing the memory and communication overhead of the program.

For large $N_\text{elec}$, the practice of deterministically promoting all high-body contributions from low-body excitations becomes computationally infeasible, and so a process by which the number of promotions can be reduced is needed.
Investigations are ongoing into efficient ways to deal with this promotion bottleneck due to the high communication cost of repeated index high-rank RDM contributions. These include a further stochastisation in which elements of the expensive loops are randomly selected, giving the largest number of realised promotions to the contributions with largest values of $\sqrt{|C_\bfi^*C_\bfj|}$.
The current implementation however undergoes deterministic promotion of all low-rank excitations to $\Gamma_{(3)}$ and contracted $\Gamma_{(4)}$ elements without further approximation.

\subsection{Stochastic estimation of MRPT intermediates}
The estimation of MRPT2 energies consists of the contraction of high rank tensors down to a scalar result.
Hence, striking the optimal balance between the degree of on-the-fly contraction and memory usage is a central concern.
Storing the full $\Gamma_{(4)}$ becomes impractical due to its $\mathcal{O}\sqrbrk{n_\text{orb}}$ scaling.
Fortunately, for NEVPT2 only two intermediates $\overline{G^A}$ and $\overline{G^B}$, formally requiring $\mathcal{O}\sqrbrk{n_\text{orb}{}^6}$ memory (in a stochastically sparse and distributed form) are required in order to estimate the sc-NEVPT2 energy, as defined in Eqs.~\ref{eq:nevpt2_intermediate_def_1} and \ref{eq:nevpt2_intermediate_def_2}.
As shown in Appendix~\ref{app:intermediate_equivalence} however, the high rank normal ordered part of the two intermediates are transpositionally equivalent, so only a single contraction denoted $\overline{G^1}$ need be sampled on the fly.
$\overline{G^A}$ and $\overline{G^B}$ can then be constructed in postprocessing by transposition of indices and contraction of ERIs with the accumulated $\Gamma^{(3)}$, as given in the Appendix.

RDMs are stored in spin-resolved, normal-ordered form, and the spin tracing is done at the end of the calculation. This scheme fully exploits permutational symmetry among creation and annihilation operators.
Handling spin when estimating the intermediate calls for a slightly different approach.
The initiator FCIQMC algorithm is generally not memory bound, but with such large distributed tensorial quantities necessary for stochastic MRPT2, the limited available memory does indeed become an important consideration, and one in which an awareness of the typical specifications of modern HPC resources becomes important.
The implemented contraction scheme is a compromise between accumulating the full, spin-resolved $\Gamma^{(4)}$, and the fully contracted and spin-averaged intermediate, $\overline{G^1}_{a' b' c', a b c}$. We instead define a new \emph{partially spin-traced} NEVPT2 intermediate, as
\begin{align}
	{G^1}&_{a'_\sigma b'_\tau c'_\sigma, a b_\tau c} \equiv \sum_{def}^\text{CAS} \sum_{\mu\nu} \\
	&\langle de | fa \rangle 
\langle \Psi_0 | 
\hat{a}_{c'_\sigma}^\dagger 
\hat{a}_{b'_\tau}^\dagger 
\hat{a}_{d_\mu}^\dagger 
\hat{a}_{e_\nu}^\dagger 
\hat{a}_{c_\nu}
\hat{a}_{f_\mu} 
\hat{a}_{b_\tau} 
\hat{a}_{a'_\sigma} 
| \Psi_0 \rangle,
\end{align}
from which the spin-free NEVPT2 intermediate can be found as
\begin{equation}
\overline{G^1}_{a' b' c', a b c}
= \sum_{\sigma\tau\nu} {G^1}_{a'_\sigma b'_\tau c'_\sigma, a b_\tau c}.
	\label{eq:spin_resolved_to_spin_traced_g1}
\end{equation}
Tracing over only two spin indices on the fly was found to be much more efficient than simply storing the spin-free intermediate on the fly i.e. tracing over all four spin indices.

Filling the partially spin resolved $G^1$ array on the fly requires more complicated logic than simply filling the fully spin traced $\overline{G}^1$.
To deal with this efficiently, the contractions and their parities were enumerated in a procedurally generated routine which given a set of four creation and four annihilation spin-operators, can be used to cycle through all orderings of free indices and contraction indices which contribute to $G^1$ and indicate when the sign of the contribution should be inverted due to the parity of the rearrangement.
By this procedure it is ensured that $\mu$ and $\nu$ are properly traced out on the fly---leaving the bulk of the spin tracing of Eq.~\ref{eq:spin_resolved_to_spin_traced_g1} to be completed at the end of the sampling.
An example of this on-the-fly promotion procedure is given in Appendix~\ref{app:nevpt2_contract}.

The CASPT2 case in the pseudocanonical basis is much simpler, since the contraction is done with the diagonal of the generalised Fock matrix, a vector quantity as opposed to the four-indexed ERIs in the NEVPT2 case.
For CASPT2, the promotion of one-, two-, and three-body excitations to $\Gamma_{(4)}$ contributions is again intercepted at the point where they would be communicated to the process storing the spin-resolved $\Gamma_{(4)}$ element, but instead the common pairs between the creation and annihilation operators of the $\Gamma_{(4)}$ element are iterated over, and the communicated elements are those of the spin-resolved, Fock-contracted intermediate, $\Gamma_{(4)}F$, as defined in Eq.~\ref{eq:CASPT2_interm}. Spin tracing over all three remaining spin indices is reserved till the end of sampling.

\section{Results}
\label{sec:results}
The stochastic MRPT2 methods developed are now applied, in order to assess performance in terms of stability, accuracy, and practicality with respect to active space size.
%These results are intended to show the capabilities of the numerical approach, rather than to produce results which can be quantitively compared with other methods, or experimental data; this will be the subject of future work.
\subsection{Chromium dimer}
Before tackling MRPT2 problems involving CAS spaces exceeding capabilities of exact methods, we first consider the systematic and random errors in the stochastic estimations compared to exact results, and determine the extent to which they can be minimised.
In this subsection, the ground state of the chromium dimer at equilibrium bond length is tackled, with a 12 spatial orbital, 12 electron CASSCF space constructed by RHF in a cc-pVTZ basis set---a challenging problem in its own right with both static and dynamic correlation contributions, but where exact determinanistic calculations can still be performed. All FCIQMC calculations are done with an initiator threshold of 3. 
All RDM and MRPT2 intermediate accumulation is done by first growing the walker population to a preset threshold value. The population is then subject to increased walker death as the shift $S_0$ is varied.
After a short period of equilibration, the 20 most highly occupied determinants are selected to define the deterministic space of the semi-stochastic adaptation.
Stochastic RDM and intermediate estimations are processed to yield MRPT2 energies every 100 iterations.
Energy errors are given as the signed difference between the stochastic estimate and the deterministic value for each of the MRPT2 energies.
\begin{figure}[!h]
	\centering
	\includegraphics[width=0.5\textwidth]{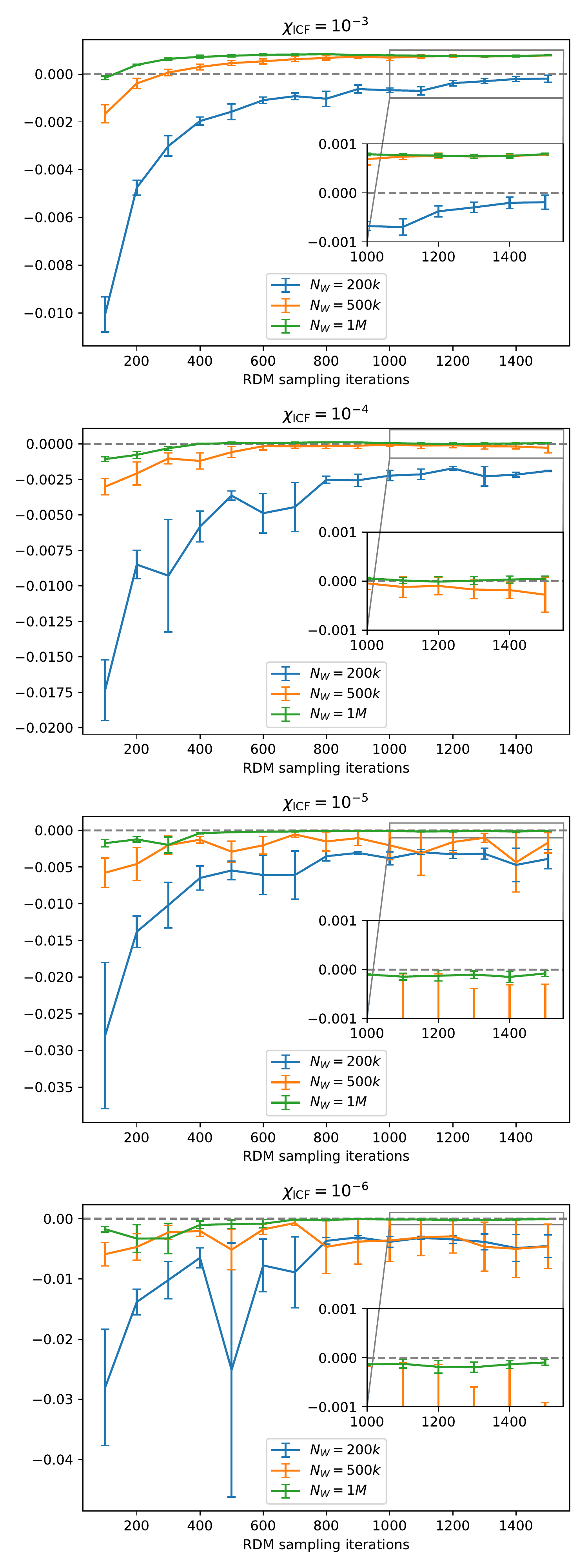}
		\caption{
			CASPT2 energy error ($\Eh$) against RDM sampling iteration number for a selection of ICF overlap thresholds and total walker numbers.
		The error bars show the standard error in the mean over 4 statistically independent runs.
		\label{fig:caspt2_vs_niter}
	}
\end{figure}

\subsubsection{CASPT2}
Figure~\ref{fig:caspt2_vs_niter} shows the accuracy of FCIQMC CASPT2 for several walker numbers and ICF overlap theshold magnitudes denoted $\chi_\text{ICF}$.
An ICF is removed from the FOIS if it corresponds to a singular value of the overlap matrix with magnitude less than $\chi_\text{ICF}$.
The system is small enough that the CASSCF optimisation can be comfortably performed by exact methods before the accumulation of RDMs and intermediates is undertaken with FCIQMC.
It is noted that for large $\chi_\text{ICF}$ ($\sim10^{-3}$), the CASPT2 energies exhibit a smooth convergence toward an incorrect value, due to systematic error introduced by the severe depletion of the FOIS dimensionality.
For all smaller thresholds, 1M walkers is sufficient to resolve CASPT2 energies accurately for even a small number of sampling iterations, while the 500k walker calculations begin to show signs of ill-conditioning in the orthogonalisation procedure of Eq.~\ref{eq:caspt2_lowdin}, due to the random error in the $\Gamma^{(3)}$ values.

However, it appears that low-systematic error results are attainable through the systematic reduction of random error resulting from increasing walker number
This confirms the predictions of the toy model in \ref{sec:conditioning}, and gives cause for optimism that the stochastic CASPT2 will acheive similar levels of accuracy in larger CAS calculations.

\subsubsection{NEVPT2}
\begin{figure}[h]
	\includegraphics[width=0.5\textwidth]{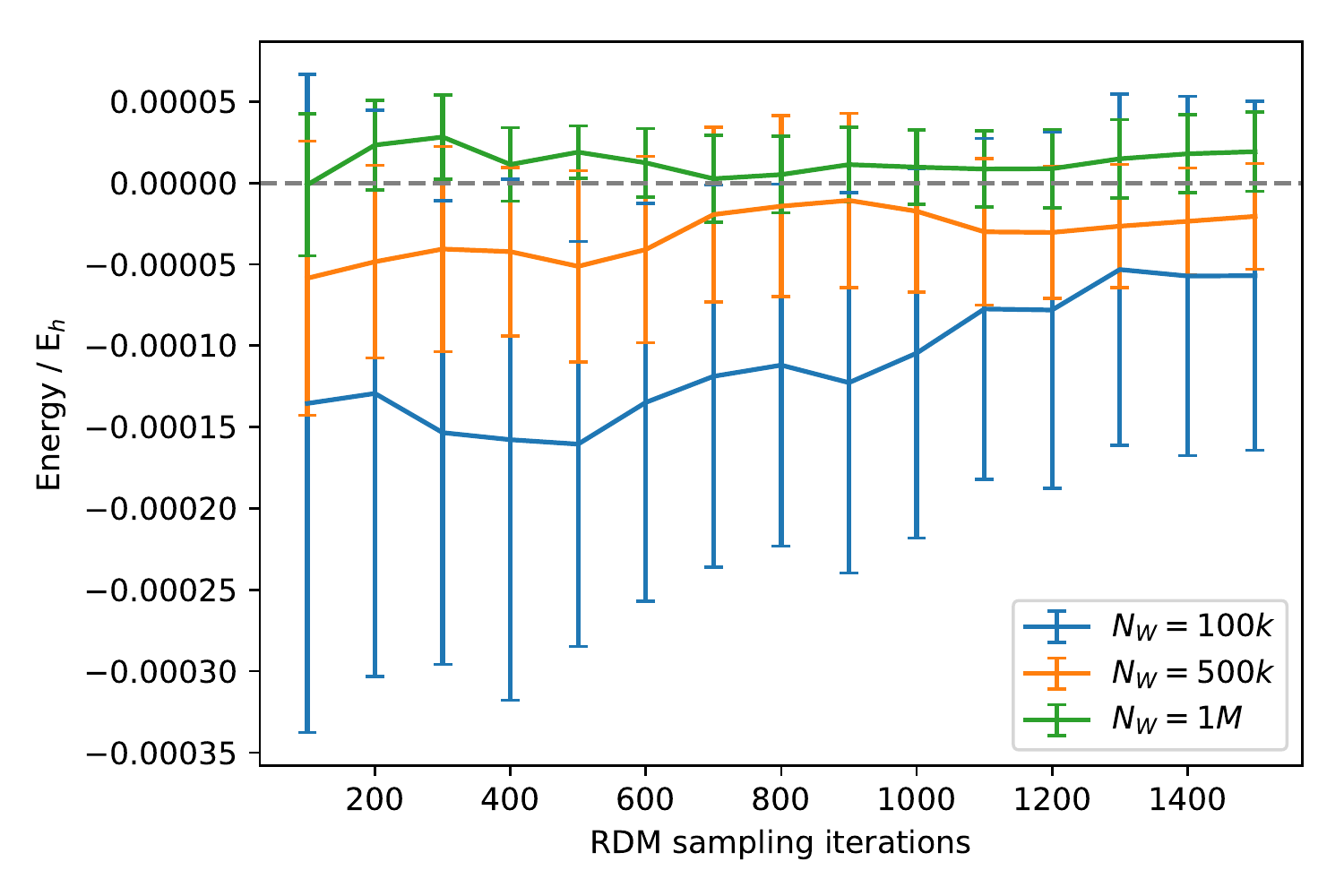}
		\caption{
			NEVPT2 energy error ($\Eh$) against RDM sampling iteration number for a selection of total walker numbers.
		The error bars show the standard error in the mean over 4 statistically independent runs.
		\label{fig:nevpt2_vs_niter}
		}
\end{figure}

The overall trend in Figure \ref{fig:nevpt2_vs_niter} shows the systematic improvability of the NEVPT2 energy estimate as the number of FCIQMC walkers is increased, in both the magnitude of the random and systematic errors.
Relative to the equivalent results for CASPT2 (Fig.~\ref{fig:caspt2_vs_niter}), we find far smaller random error bars, and much higher accuracy achieved overall for the same number of walkers. This agrees with the rationalization in~\ref{sec:conditioning}.
From this study it seems that longer sampling periods with a low-fidelity wavefunction will fail to resolve the NEVPT2 energy more precisely and accurately than a large walker population for a short sampling duration.
It is notable that stochastic NEVPT2 is more accurate and with smaller errors than the equivalent CASPT2 calculations, which is again anticipated by the conditioning experiment.
These results are a strong indication that NEVPT2 is a good candidate for MRPT2 stochastisation for system sizes beyond deterministically tractable limits, with 1M walkers easily achieving $\mathcal{O}[10^{-4}]\mathrm {E_h}$ accuracy.

\subsection{Application to Cu${}_2$O${}_2$(NH${}_3$)${}_2{}^{2+}$}
A well known bonding motif in biologically-relevant molecules relates to the peroxo and bis-($\mu$-oxo) isomers of a Cu${}_2$O${}_2^{2+}$ core, corresponding to the II and III oxidation states of the copper atoms respectively.
It is found in a variety of different metalloproteins, such as haemocyanin and other enzymes such as catechol oxidase and tyrosinase \cite{Werner_cu2o2}.
Such macromolecules are much too large to be practically studied with high-accuracy post-HF methods, however electron correlation effects are known to play an important role in determining the energetics and other properties of the Cu${}_2$O${}_2^{2+}$ core.
Hence it is common to study model coordination complexes, in which the protein ligands are replaced by simpler fragments, with quantum many-body methods.
Following Werner\cite{Werner_cu2o2} and Pierloot \cite{doi:10.1021/acs.jctc.6b00714}, we choose to study the peroxo and bis-($\mu$-oxo) isomers of the Cu${}_2$O${}_2$(NH${}_3$)${}_2{}^{2+}$ model system, where we use a cc-pVTZ basis on all Cu atoms and cc-pVDZ basis for other elements, giving a total of 222 orbitals and 92 correlated electrons with a (24,24) active space. The geometry is taken from Ref.~\cite{doi:10.1021/acs.jctc.6b00714}, and we use a HF basis. Combining the effects of NEVPT2 with a prior stochastic CASSCF orbital optimization is possible, and will be considered in the future\cite{doi:10.1021/acs.jctc.5b00917,doi:10.1021/acs.jctc.5b01190}.

\begin{figure}[h]
	\centering
	\includegraphics[width=0.5\textwidth]{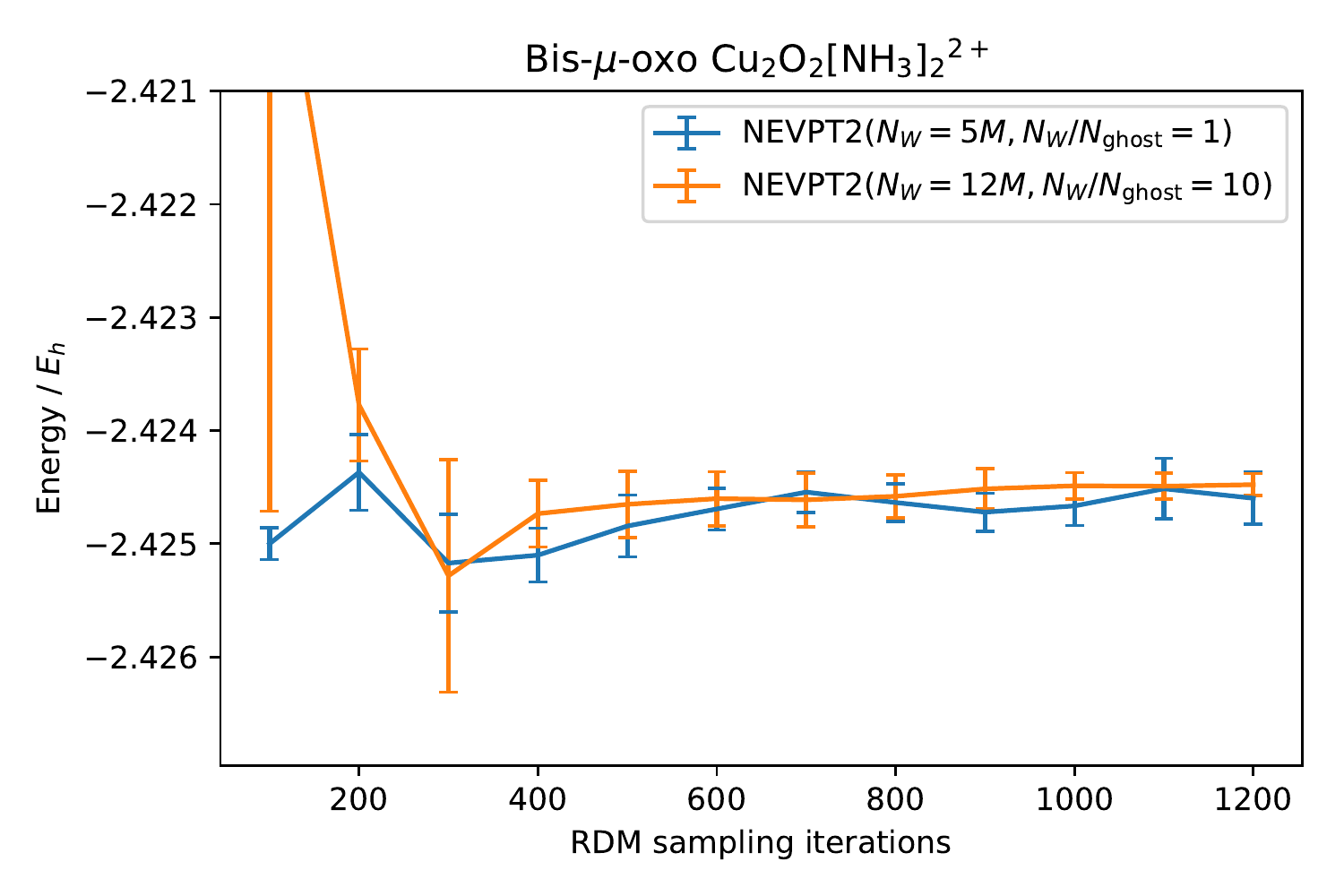}
	\caption{
		Total NEVPT2 energy ($\Eh$) against RDM sampling iterations for two different FCIQMC walker numbers ($N_w$) and two different coarseness of the sampling of higher-body excitations in the active space ($N_w/N_{\textrm{ghost}}$ which denotes the number of higher-body excitations compared to a standard (up to two-body) FCIQMC sampling attempt). The system is in the bis-$\mu$-oxo isomer, with a (24,24) active space.
		The error bars show the standard error in the mean over four statistically independent runs.
		\label{fig:bis_mu_oxo_nevpt2}
	}
\end{figure}

\begin{figure}[h]
	\centering
	\includegraphics[width=0.5\textwidth]{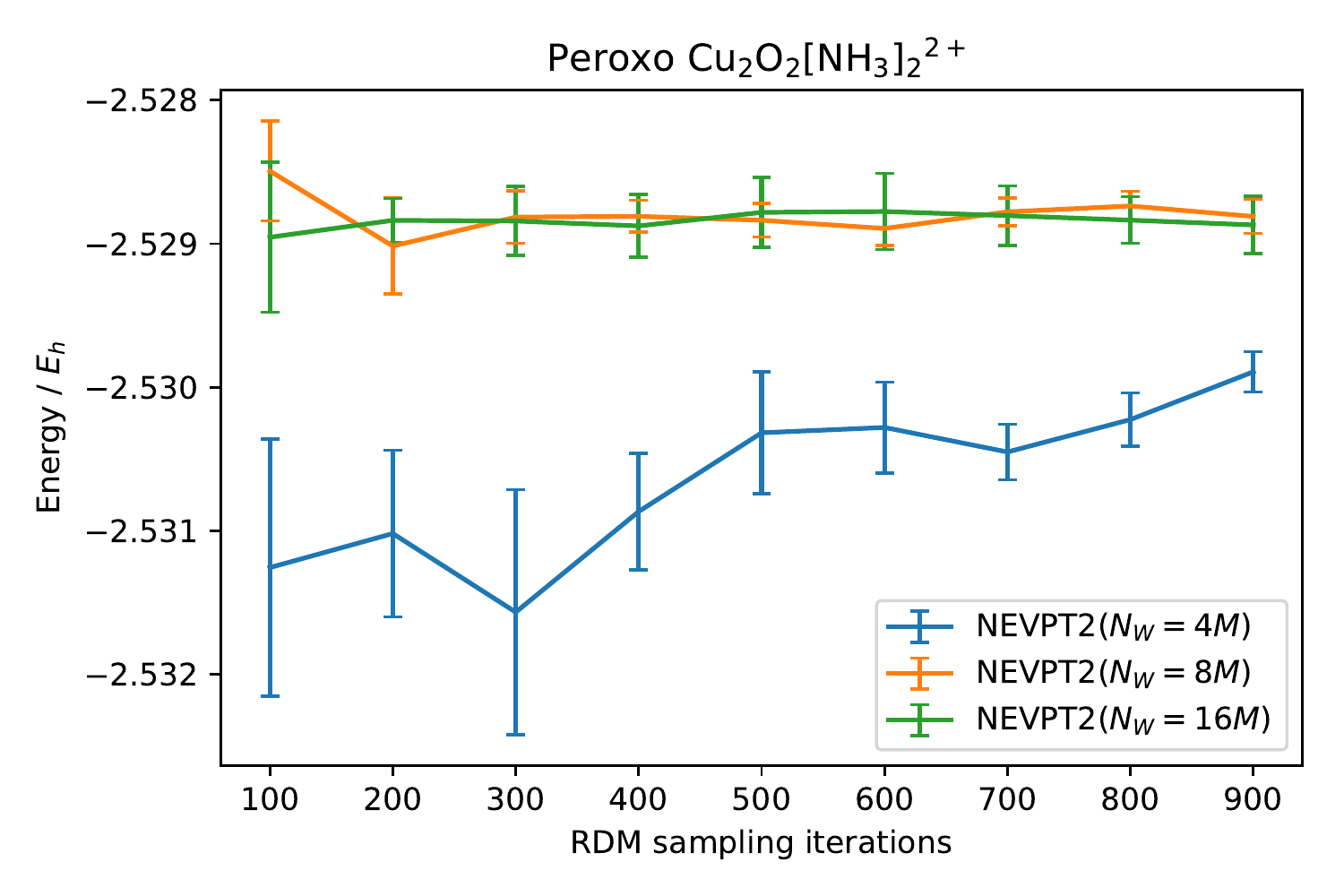}
	\caption{
		Total NEVPT2 energy ($\Eh$) for the peroxo isomer with a (24,24) active space against RDM sampling iterations for a selection of total walker numbers, whilst keeping the high-body sampling fixed ($N_w/N_{\textrm{ghost}}=1$).
		The error bars show the standard error in the mean over four statistically independent runs, with the deterministic subspace ($N_{SS}$) given by 500 determinants.
		\label{fig:peroxo_nevpt2}
	}
\end{figure}

Application of FCIQMC--CASPT2 to these systems suffered from large random errors, even with $\chi_\text{ICF}=10^{-3}$, at which the energy correction would be subject to a non-negligible systematic error even if the effects of the numerical instability in the ICF overlap could be mitigated. Therefore, for the present work, we focus on FCIQMC--NEVPT2, which has been shown to exhibit far less sensitivity to the effects of random noise in its evaluation.
%The effort to apply FCIQMC--CASPT2 to these systems was met with little success, with large random errors in the CASPT2 energy evident even with $\chi_\text{ICF}=10^{-3}$, at which the energy correction would be subject to a large systematic error even if the effects of the stochastic estimation of $\overline{\Gamma}^{(3)}$ could be mitigated.
%At the present time, we focus on FCIQMC--NEVPT2, which was shown in the previous subsection to be less sensitive to noise in the RDM estimates.
Figure \ref{fig:bis_mu_oxo_nevpt2} shows the convergence with respect to RDM sampling iterations for a (24, 24) CAS of the bis($\mu$-oxo) core geometry for two different sets of FCIQMC-NEVPT2 parameters.
Complete active spaces of this dimension are beyond the scope of traditional active space solvers, let alone NEVPT2 implementations.

Both calculations were performed with 5,000 determinants in the deterministic space of the semi-stochastic adaptation, but the 12M walker calculations used a coarser sampling of the active space three- and four-body excitations.
The 5M walker calculation attempted a higher-body ghost spawn for every walker each iteration ($N_w/N_{\textrm{ghost}}=1$), whereas the 12M walker calculation attempted to spawn one ghost walker only every 10 walkers on average.
The effect of the coarse sampling is initially stark, with the first few sampling periods of the high-body quantities corresponding to very large statistical error in the NEVPT2 energy for the 12M walker calculations.
However, by the time 1,200 RDM sampling iterations have elapsed, the sampling quality of these contributions is sufficient for both granularities of ghost spawns, and the errors become determined by the accuracy with which the wavefunction is sampled. The preferred simulation is therefore to run with a larger walker population, and sparse sampling of the higher-body excitations, which correspond predominantly to the direct contributions to the higher-body intermediates, where all indices are unique. We want to consider further the required walker population to faithfully sample these intermediates and extract accurate NEVPT2 energies, as the number of walkers will ultimately correspond to the memory requirements of the calculation.
%Hence, in the long run, the errors for the larger walker population are much smaller.

For this, we consider the peroxo isomer, with Figure \ref{fig:peroxo_nevpt2} showing a set of results where coarseness the higher-body sampling is kept constant ($N_w/N_{\textrm{ghost}}=1$), and showing the convergence for three different walker numbers, at 4M, 8M and 16M. 
The NEVPT2 energy is converged to well within chemical accuracy for the 8M and 16M walker calculations, but 4M walkers provides an inadequate representation of the active space wavefunction for NEVPT2 estimates to contain low enough random and initiator error for the accurate energy.
We can contrast this required number of walkers to converge the standard CASCI projected energy (taken with a multiconfigurational trial wave function of 200 determinants). For this, the 4M walker projected energy is converged within errorbars with the 8M walker estimate, indicating that the initiator error at this walker number is sufficient to gain an accurate CASCI energy, but not NEVPT2 energy. This also agrees with the loose rule-of-thumb, which asserts that walker populations with at least 50,000 walkers on the RHF determinant are sufficient to converge the initiator error (for the projected energy) to well within chemical accuracy -- however demonstrating explicit convergence with respect to walker number is always preferable to relying on this. For this system, 4M walkers gives 61,000 walkers on the RHF determinant, agreeing with this rule of thumb, but to converge the initiator error for the NEVPT2 energy requires approximately double this number of walkers. 
It does not seem unreasonable that the NEVPT2 energy is more sensitive to the number of walkers sampling the underlying wave function and the fidelity of its representation, as it depends on the relatively populations between highly excited configurations (whereas the original Hamiltonian for the energy is a more local operator). %This is likely true even after the initiator error has been saturated from the projected energy FCIQMC-CASCI estimator.
%The 200 determinant trial wavefunction energy is converged for all walker numbers on the plot, with the 4M walker mean trial-projected energy matching that of the 8M walker estimate within error bars for each independent run.
%The steady walker populations on the RHF determinant are roughly 61k, 117k, and 226k for the 4M, 8M, and 16M walker calculations respectively.
%It is an established rule of thumb that 50k walkers on the HF determinant is sufficient to saturate initiator error for $n_\text{add}=3$, although initiator curves should always be produced to ensure convergence within stochastic error bars.
%Nevertheless, this observation is often a useful guide, and it is noted that convergence of the strongly-contracted NEVPT2 energy with respect to walker number within error bars has required approximately twice the walker population on the HF determinant than the trial wavefunction energy.

\begin{figure}[h]
	\centering
	\includegraphics[width=0.5\textwidth]{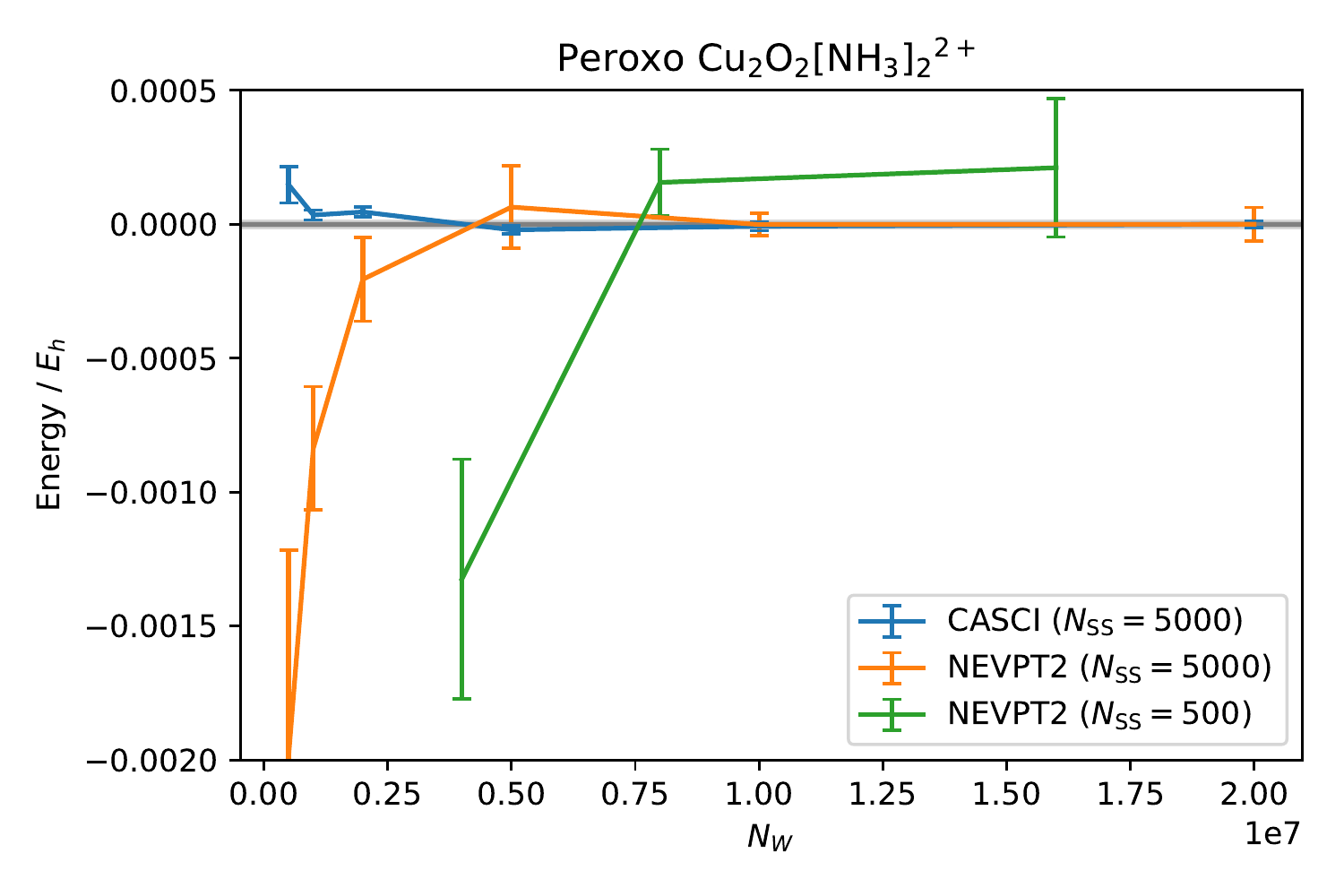}
	\caption{
		FCIQMC-NEVPT2 and CASCI energy convergence as a function of walker number for the peroxo isomer with a (24, 24) CAS, additionally showing the effect of a ten-fold increase in the number of determinants in the deterministic space.
		The energies are given relative to the 20M walker best values, at which they are evidently converged well beyond the standard definition of chemical accuracy ($\sim1$~m$\Eh$) for a deterministic space of 5,000 determinants. All RDMs are accumulated over 500 MC cycles.
		\label{fig:casci_nevpt2_vs_nw}
	}
\end{figure}

The resolution of the sampling of the NEVPT2 intermediates can be goverened by essentially three paramters: the number of walkers controlling the resolution of the wave function amplitudes and saturation of any initiator error ($N_w$), the coarseness of the higher-body sampling (goverened by $N_w/N_{\textrm{ghost}}$), and additionally the size of the deterministic subspace of the simulation ($N_\text{SS}$). This final value is important, since all higher-body sampling and intermediates between the deterministic subspace determinants are included exactly within the required estimates. 
To try to disentagle these effects, in Fig.~\ref{fig:casci_nevpt2_vs_nw} we consider the convergence of the FCIQMC-NEVPT2 and FCIQMC-CASCI energy for a (24,24) active space with respect to walkers for two different deterministic subspace sizes.
%The number of walkers in the system, $N_W$ governs the extent to which initiator error has been saturated, and hence the convergence of the average walker distribution to the exact ground state eigenvector of the Hamiltonian.
%As remarked in subsection \ref{rdm_fciqmc}, once the stochastic evolution of the wavefunction in both replicas reaches a high degree of fidelity with the sampling, 
% respect to the exact eigenvector of $H$, the systematic improvement of the RDMs can proceed in a number of ways.
%Generally, it seems that singles and doubles are generated too infrequently for an effective sampling of the higher-body RDMs when only a single spawning attempt is made per per unit walker.
%This is intuitive, since the promotion procedure essentially magnifies the sparseness of the low excitation level sampling.
%Figure \ref{fig:casci_nevpt2_vs_nw} shows the convergence of the NEVPT2 energy for a (24, 24) active space with respect to walker number, and demonstrates the interplay between two methods of improving the estimation of the higher-body RDMs.
By around 5M walkers, the walker distribution resolves the exact wavefunction well enough to converge the initiator error for the CASCI variational energy $\text{Tr}\sqrbrk{{\Gamma}_{(2)}H}$ to within small random error bars, as argued previously.
This provides a good indication that with 5M walkers, the dominant issue is the required sampling of the NEVPT2 intermediates.
The plot shows that increasing the size of the deterministic space is an effective way to avoid the greater additional expense of increasing $N_W$.
With $N_\text{SS}=5,000$, the NEVPT2 energy at 5M walkers is more precise and more accurate than the equivalent calculation with $N_\text{SS}=500$ and a three-fold increase in $N_w$. While the random errors are larger than the equivalent $N_w$ FCIQMC-CASCI calculation, the rate of convergence of initiator error in the NEVPT2 is now broadly similar to the CASCI, indicating that much of the error was previously coming from incomplete sampling of the intermediates. More applications are needed before a clearer picture on the walker numbers required for the MRPT2 calculations can be confirmed, and its scaling with active space size, but these results indicate that if an appropriate sampling of the higher-body intermediates is possible, that the number of walkers required is certainly of the same order as that required to sample the underlying FCIQMC wave function. This brings into reach accurate quantum chemical treatments of a large array of multiconfigurational systems which are currently limited by the size of the CAS practicable in deterministic algorithms.

\begin{figure}[h]
	\centering
	\includegraphics[width=0.5\textwidth]{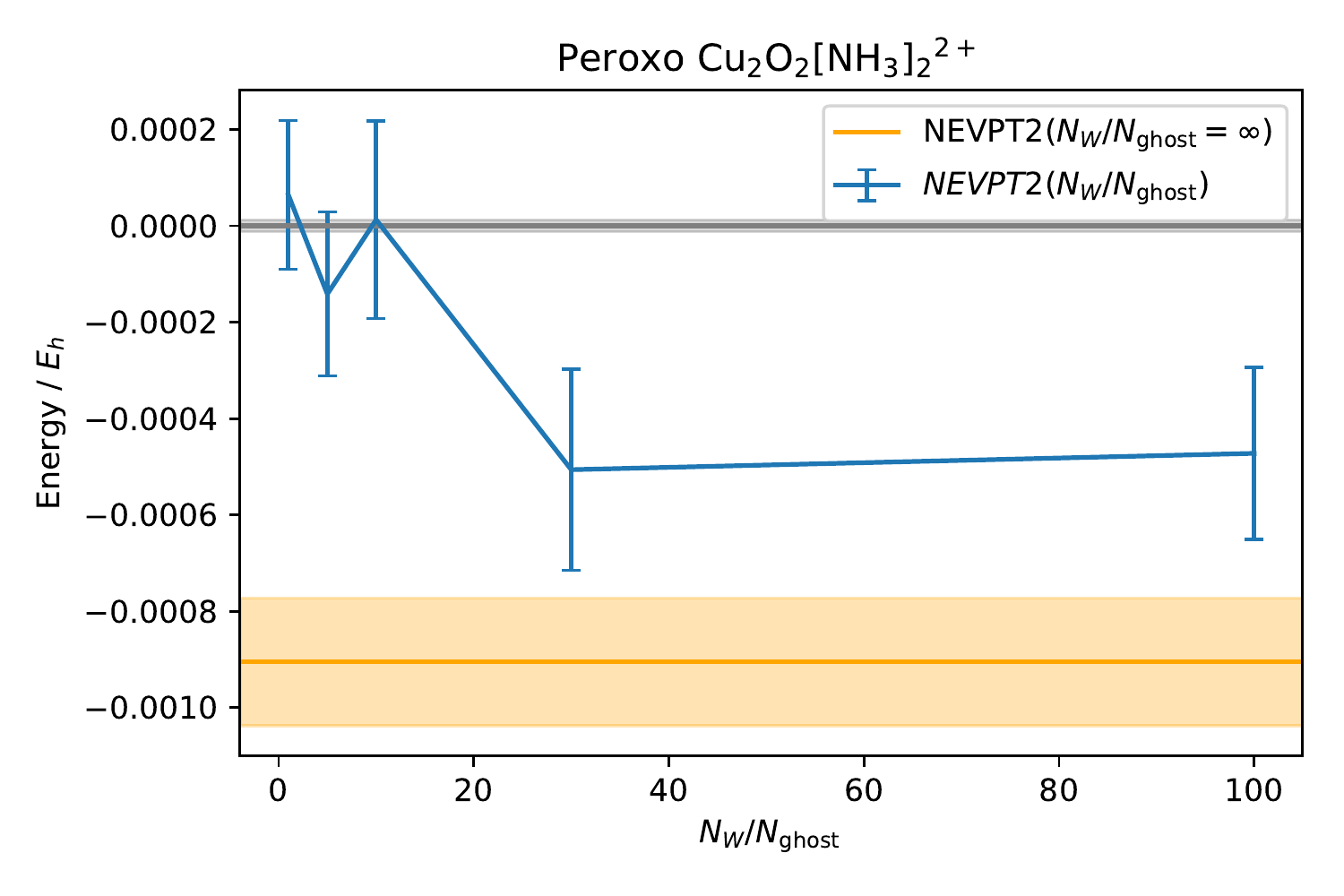}
	\caption{
		FCIQMC-NEVPT2 energy convergence as a function of the number of ghost spawnings attempted per unit walker for the peroxo isomer with a (24, 24) CAS, with 5,000 deterministic determinants and 5M walkers. 
		The energies are given relative to the converged 20M walker values. The $N_w/N_{\textrm{ghost}}=\infty$ line denotes a simulation where the only three- and four-body sampling to the required higher-body intermediates derive from the deterministic sampling within the 5,000 core determinants and all stochastic higher-body sampling is supressed.
		\label{fig:nevpt2_energy_vs_HB_granularity}
	}
\end{figure}

Another factor which can be varied to improve sampling of ${\Gamma}_{(3)}$ and ${\Gamma}_{(4)}$ (or rather the intermediates sampled from these) is the granularity with which the three- and four-body excitations are sampled in the active space. 
Figure \ref{fig:nevpt2_energy_vs_HB_granularity} investigates the effect of decreasing the frequency at which ghost walkers are generated to sample these quantities, while holding $N_w$ constant at 5M, and $N_\text{SS}$ at 5,000.
For $N_w/N_\mathrm{ghost}\leq10$, the sampling frequency of such contributions to the higher-body effective quantities is high enough to ensure that sparsity of three- and four-body excitation sampling is not the dominant source of error.
However, with $N_w/N_\mathrm{ghost}=30$ and 100, the performance penalty of a low frequency of ghost spawning becomes obvious. 
Also included in the plot is the $N_W/N_\mathrm{ghost}=\infty$ and its error bar, showing clearly the detriment of totally neglecting stochastic three- and four-body exctiations.
It is important to note that at $N_w/N_\mathrm{ghost}=\infty$, the three- and four-body connections \emph{within} the deterministic space are not neglected, and so the contributions within these 5,000 determinants is still included exactly. This demonstrates that the sparse stochastic sampling afforded by the random higher-body excitation generation and ghost spawning within the stochastic space of all the remaining determinants still contributes a noticeable and important effect on the overall accuracy of the method.

Combining the converged results of these systems, we find that the FCIQMC-CASCI(24,24) estimates of the energy difference between the isomers result in the bis-$\mu$-oxo isomer being more stable than the peroxo isomer by 8.8mE$_{\textrm h}$. However, the inclusion of the FCIQMC-NEVPT2 correlation of the external space reverses the qualitative sense of the CASCI result. The walker converged FCIQMC-NEVPT2 estimates of the (24,24) system give a total energy of -3541.9390(1) for the bis-$\mu$-oxo isomer, while the peroxo isomer is -3542.0345(2), resulting in the peroxo isomer being more stable by a significant 95.6(2)mE$_{\textrm h}$.
%The energy difference is the main observable of interest for this system.
%For the bis($\mu$-oxo) core, the 12M walker trial wavefunction (CASCI) energy is $-3539.514480~\Eh$. The FCIQMC--NEVPT2 energy at 1,200 RDM sampling iterations was found to be $-2.424478~\Eh$ with an error of 0.1~m$\Eh$.
%For the peroxo core on the other hand, the 16M walker trial wavefunction (CASCI) energy is -3539.505675$~\Eh$. The FCIQMC--NEVPT2 energy at 900 RDM sampling iterations was found to be $-2.528869~\Eh$ with an error of 0.2~m$\Eh$.
%The MRPT2 correction to the CASCI energy is the dominant source of error for the large walker populations, and so errors are not quoted for the CASCI energies, which are assumed to be signficantly more finely resolved.
%Thus, the bis($\mu$-oxo) geometry exceeds the peroxo geometry in energy by approximately 0.0956$~\Eh$, reversing the qualitative sense of the CASCI result.
Pierloot et al \cite{doi:10.1021/acs.jctc.6b00714}, and Rode \& Werner \cite{Werner_cu2o2} agree with these conclustions that the peroxo geometry is more stable, with the former evaluating the cumulant-approximated CASPT2 energy difference with DMRG at 36$~\mEh$ in their own (24, 24) CAS calculation; and the latter remarking that this finding is in qualitative agreement with the experimental observation that the peroxo form occurs in enzymes.
Closer agreement in energy with these references is not expected given the difference in MRPT2 methods applied, and the differing basis sets and active space selection employed.
It is interesting however---and affirming of the importance of accurately treating dynamic correlation---that seemingly the correct identification of peroxo as the more stable core arises only after a treatment of the problem by FCIQMC--NEVPT2.

\section{Conclusion}
In this work, FCIQMC has been demonstrated to be a capable and efficient active space solver for the computation of fully internally-contracted multireference perturbation theory corrections to molecular electronic energies.
NEVPT2 energies were estimated to high accuracy for a (24, 24) CAS of a strongly correlated transition metal complex beyond that capable by traditional means.
CASPT2 calculations were found to be less amenable to a similar stochastic treatment, as the inversion of RDM estimates proved to be be highly sensitive to random errors introduced in the formulation.
However, the results obtained in this study give cause for optimism that FCIQMC--NEVPT2 can be applied---with some algorithmic refinements---to extremely large active space correlation problems. By introducing additional means to control the effective sampling of the higher-body terms, the convergence of the FCIQMC-NEVPT2 energies were found to require a similar number of walkers as the faithful sampling of the active space wave function required for FCIQMC-CASCI.

This study has shown that the promotion of low excitation level connections between determinants and the generation of non-propagating higher-body excitations are both important classes of contributions to the required intermediates, and that on-the-fly contraction of $\mathcal{O}\sqrbrk{n_\text{orb}{}^6}$-scaling MRPT2 intermediates is an effective way to ensure sampling of the full four-body RDM without incurring its $\mathcal{O}\sqrbrk{n_\text{orb}{}^8}$ memory cost. Furthermore, efficient distributed and sparse storage of these quantities is natural in the FCIQMC framework.
The most immediate target for future improvement efforts is the bottleneck introduced by the need for promotions.
Further stochastisation of these contributions and importance-sampled contraction of the intermediates based on the Fock matrix (CASPT2) or ERIs (NEVPT2) could improve the efficiency of higher-body RDM estimation.

%Further refinement of the computational approach, particularly with regard to storage and communication of multidimensional arrays, will be the focus of future development efforts.
%While steps have already been taken to manage the cost of accumulating the RDMs and MRPT2 intermediates on the fly, such as private storage of the diagonal elements of $\Gamma_{(3)}$ and $\Gamma_{(4)}$, a more sophisticated partitioning between distributed and privately accumulated elements could have a marked positive effect on the efficiency of the algorithm.
%This would entail accumulating certain frequently sampled nondiagonal elements of the high-rank RDMs privately then aggregating over all processes at the end of the sampling run.

%An even better approach would be to introduce the option of accumulating RDMs in a hybrid parallelisation scheme instead of the purely distributed system used currently; but this would require substantial changes to the existing codebase. 

\section{Acknowledgements}

G.H.B. gratefully acknowledges support from the Royal Society via a University Research Fellowship, as well as funding from the European Union's Horizon 2020 research and innovation programme under grant agreement No. 759063. We are also grateful to the UK Materials and Molecular Modelling Hub for computational resources, which is partially funded by EPSRC (EP/P020194/1).

\providecommand{\latin}[1]{#1}
\providecommand*\mcitethebibliography{\thebibliography}
\csname @ifundefined\endcsname{endmcitethebibliography}
  {\let\endmcitethebibliography\endthebibliography}{}

\appendix

\section{RDM definitions}
\label{app:rdm}
Numerous different notations for RDMs appear in the literature.
In this work the following notational conventions are adopted.

An $n$-RDM whose elements correspond to expectation values of normal ordered products of second quantised operators are always denoted with subscript indices and with a comma delimiting hole (creation) and electron (annihilation) operator indices:
\begin{equation}
	\Gamma_{i_1 i_2 ... i_n, j_1 j_2 ... j_n} \equiv
	\langle \Psi |
	\hat{a}_{i_1}^\dagger \hat{a}_{i_2}^\dagger ... \hat{a}_{i_n}^\dagger  
	\hat{a}_{i_n} \hat{a}_{i_{n-1}} ... \hat{a}_{i_1}
	| \Psi \rangle.
\end{equation}
The RDM elements corresponding to expectation values of products of single excitation operators are denoted with superscript creation operator indices and subscript annihilation operator indices:
\begin{equation}
	\Gamma^{i_1 i_2 ... i_n}_{j_1 j_2 ... j_n} \equiv
	\langle\Psi|
	\hat{a}_{i_1}^\dagger \hat{a}_{j_1}
	\hat{a}_{i_2}^\dagger \hat{a}_{j_2} ... 
	\hat{a}_{i_n}^\dagger \hat{a}_{j_n}
	|\Psi\rangle.
\end{equation}
Spin-free RDMs are denoted with an overbar:
\begin{multline}
	\overline{\Gamma}_{p_1 p_2 ... p_n, q_1 q_2 ... q_n} \equiv \\
	\sum_{\sigma_1 \sigma_2 ... \sigma_n}
	\langle\Psi|
	\hat{a}_{p_1\sigma_1}^\dagger \hat{a}_{p_2\sigma_2}^\dagger ... \hat{a}_{p_n\sigma_n}^\dagger  
	\hat{a}_{q_n\sigma_n} \hat{a}_{q_{n-1}\sigma_{n-1}} ... \hat{a}_{q_1\sigma_1}
	|\Psi\rangle;
\end{multline}
\begin{multline}
	\overline{\Gamma}^{p_1 p_2 ... p_n}_{q_1 q_2 ... q_n} \equiv
	\langle\Psi|
	\excit{p_1}{q_1}
	\excit{p_2}{q_2}...
	\excit{p_n}{q_n}
	|\Psi\rangle \equiv \\
	\sum_{\sigma_1 \sigma_2 ... \sigma_n}
	\langle\Psi|
	\hat{a}_{p_1\sigma_1}^\dagger \hat{a}_{q_1\sigma_1}
	\hat{a}_{p_2\sigma_2}^\dagger \hat{a}_{q_2\sigma_2} ... 
	\hat{a}_{p_n\sigma_n}^\dagger \hat{a}_{q_n\sigma_n}
	|\Psi\rangle.
\end{multline}
When referring to an RDM of some rank $n$ instead of a particular element, the notations $\Gamma_{(n)}$ and $\overline{\Gamma}_{(n)}$ are used to denote spin resolved and spin free RDMs respectively.

%The nomenclature \emph{excitation level} is used in reference to the number of hole orbital indices which are not also electron orbital indices. E.g. a diagonal RDM element corresponds to an excitation level of zero, while a double replacement corresponds to an excitation level of two.

\section{Expressing product of single excitation RDMs in terms of normal ordered RDMs}
The internally contracted functions of multireference perturbation theories are usually expressed as products of single excitation operators acting on the reference space. In FCIQMC, it is natural to sample the RDMs in normal-ordered form. Hence, after the RDMs and intermediates have been accumulated, the following transformations are necessary before the individual subspace energies can be estimated.

\begin{equation}
\overline{\Gamma}_{a}^{i} = \overline{\Gamma}_{i,a}
\end{equation}
\begin{equation}
\overline{\Gamma}_{ab}^{ij} = \overline{\Gamma}_{ij,ab} +\overline{\Gamma}_{i,b}\delta_{aj}
\end{equation}
\begin{align}
\begin{split}
	\overline{\Gamma}_{abc}^{ijk} = &\overline{\Gamma}_{ijk,abc} +\overline{\Gamma}_{ik,bc}\delta_{aj} +\overline{\Gamma}_{ij,ac}\delta_{bk} +\\
	&\overline{\Gamma}_{ij,cb}\delta_{ak} +\overline{\Gamma}_{i,c}\delta_{aj}\delta_{bk}
\end{split}
\end{align}
\begin{align}
\label{4POSENORD}
\begin{split}
	\overline{\Gamma}_{abcd}^{ijkl} = &\overline{\Gamma}_{ijkl,abcd} +\overline{\Gamma}_{ijk,adc}\delta_{bl} +\overline{\Gamma}_{ijl,acd}\delta_{bk} +\\
	&\overline{\Gamma}_{ikl,bcd}\delta_{aj} +\overline{\Gamma}_{ijl,cbd}\delta_{ak}+\overline{\Gamma}_{ijk,abd}\delta_{cl} +\\
	&\overline{\Gamma}_{ijk,dbc}\delta_{al} +\overline{\Gamma}_{ij,ad}\delta_{bk}\delta_{cl} +\overline{\Gamma}_{ik,dc}\delta_{aj}\delta_{bl} +\\
	&\overline{\Gamma}_{ik,bd}\delta_{aj}\delta_{cl} +\overline{\Gamma}_{ij,dc}\delta_{bk}\delta_{al} +\overline{\Gamma}_{il,cd}\delta_{aj}\delta_{bk} +\\
	&\overline{\Gamma}_{ij,db}\delta_{ak}\delta_{cl} +\overline{\Gamma}_{ij,cd}\delta_{ak}\delta_{bl} +\overline{\Gamma}_{i,d}\delta_{aj}\delta_{bk}\delta_{cl}
\end{split}
\end{align}

\section{Equivalence of normal-ordered four-RDM parts of NEVPT2 intermediates\label{app:intermediate_equivalence}}
The expectation values of Eqs.~\ref{eq:nevpt2_intermediate_def_1} and \ref{eq:nevpt2_intermediate_def_2} can be decomposed as follows:
\begin{align}
\begin{split}
	\overline{G^A}_{a' b' c', a b c} = &\sum_{def}^{N_\text{act}} \langle de | fa \rangle \overline{\Gamma}_{c'b'de, a'bfc} +\\
	&\sum_{def}^{N_\text{act}} \langle de | fa \rangle \rndbrk{\overline{\Gamma}^{c'b'de}_{a'bfc}-\overline{\Gamma}_{c'b'de, a'bfc}}\\
	&\equiv \overline{G^{A1}}_{a' b' c', a b c} + {\overline{G^{A2}}}_{a' b' c', a b c}
\end{split}
\end{align}

\begin{align}
\begin{split}
	\overline{G^B}_{a' b' c', a b c} = &\sum_{def}^{N_\text{act}} \langle dc | fe \rangle \overline{\Gamma}_{c'b'da, a'bfe} +\\
	&\sum_{def}^{N_\text{act}} \langle dc | fe \rangle \rndbrk{\overline{\Gamma}^{c'b'da}_{a'bfe}-\overline{\Gamma}_{c'b'da, a'bfe}}\\
	&\equiv \overline{G^{B1}}_{a' b' c', a b c} + {\overline{G^{B2}}}_{a' b' c', a b c}
\end{split}
\end{align}

While $\overline{G^A}$ and $\overline{G^B}$ are distinct tensors unrelated by index transposition, it can be shown that $\overline{G^{A1}}$ is transpositionally equivalent to $\overline{G^{B1}}$:

\begin{equation*}
\overline{G^{B1}}_{a' b' c', a b c} = \sum_{def}^{N_\text{act}} \langle dc | fe \rangle \overline{\Gamma}_{c'b'da, a'bfe}
\end{equation*}
switch indices:  $a\leftrightarrow c$, $a'\leftrightarrow c'$, $b'\leftrightarrow b$, $d\leftrightarrow f$
\begin{equation*}
\overline{G^{B1}}_{c' b a, a' b' c} = \sum_{def}^{N_\text{act}} \langle fa | de \rangle \overline{\Gamma}_{a'bfc, c'b'de}
\end{equation*}
and apply hermiticity symmetries to real-valued integrals and RDM elements to obtain:
\begin{align*}
	\overline{G^{B1}}_{c' b a, a' b' c} &= \sum_{def}^{N_\text{act}} \langle de | fa \rangle  \overline{\Gamma}_{c'b'de, a'bfc} \\
	&= \overline{G^{A1}}_{a' b' c', a b c}
\equiv \overline{G^{1}}_{a' b' c', a b c}.
\end{align*}

Thus, only one tensor need be estimated in order to compute an estimate of both NEVPT2 intermediates.

\section{On-the-fly contraction of NEVPT2 intermediates}
\label{app:nevpt2_contract}
Let the sampled $n$-body connection between determinants:
$$
|D_\mathbf{i}\rangle
\text{ and }
|D_\mathbf{j}\rangle =
\hat{p}^\dagger_{1\sigma_1} \hat{p}^\dagger_{2\sigma_2} ... \hat{p}^\dagger_{n\sigma_n}
\hat{q}_{n\tau_n} \hat{q}_{(n-1)\tau_{(n-1)}} ... \hat{q}_{1\tau_1}
|D_\mathbf{i}\rangle
$$
with CI coefficients 
$C_\mathbf{i} \equiv \langle \Psi | D_\mathbf{i} \rangle$ and 
$C_\mathbf{j} \equiv \langle \Psi | D_\mathbf{j} \rangle$ be denoted 
$({p_1\sigma_1}, {p_2\sigma_2}, ..., {p_n\sigma_n}) \leftarrow
({q_1\tau_1}, {q_2\tau_2}, ..., {q_n\tau_n})$, where $p_i$ and $q_i$ are the hole (creation) and electron (annihilation) spatial orbital indices, and $\sigma_i$ and $\tau_i$ are the corresponding spin functions - either $\alpha$ or $\beta$.
The \emph{spin signature} of the connection is denoted $\sigma_1\sigma_2...\sigma_n \leftarrow \tau_1\tau_2...\tau_n$
If $n<4$ then the $4-n$ remaining unspecified indices are deterministically inserted into the hole and electron index arrays.

The full connection determines:
\begin{itemize}
	\item Five free indices $b_\tau, c_\nu, a'_\sigma, b'_\tau, c'_\sigma$
	\item Three contraction indices $d_\mu, e_\nu, f_\mu$
\end{itemize}
The remaining unspecified index is $a$, which must be iterated over and contributions to all corresponding $G^1$ elements accumulated. Only the necessary spin averaging and contracting is done on the fly.

In order to correctly contribute to all $G^1$ elements, we must first anticommute the electron indices till the hole and electron parts of the overall spin signature match, then loop over all spin signature conserving choices for the operators $\hat{a}_{d_\mu}^\dagger, \hat{a}_{e_\nu}^\dagger, \hat{a}_{c_\nu},$ and $\hat{a}_{f_\mu}$.
The remaining four indices remain spin resolved and ordered only to be spin averaged at the end of the accumulation.

Suppose for example the full connection has the spin signature $\alpha\beta\alpha\alpha\leftarrow\alpha\alpha\alpha\beta$, and we assign the following labels to the hole and electron spatial indices $p_1p_2p_3p_4\leftarrow q_1q_2q_3q_4$.

The first contribuing ordering is $p_1p_2p_3p_4 \leftarrow q_1q_4q_2q_3$ which is parity negative. This ordering makes the following identifications for the spin, free and contraction indices:
$$
\sigma \rightarrow \alpha, \,
\tau \rightarrow \beta, \,
\mu \rightarrow \alpha, \,
\nu \rightarrow \alpha;
$$
$$
a' \rightarrow q_3, \,
b' \rightarrow p_2, \,
c' \rightarrow p_1, \,
b \rightarrow q_2, \,
c \rightarrow q_1;
$$
$$
d \rightarrow p_3, \,
e \rightarrow p_4, \,
f \rightarrow q_4;
$$

and therefore makes the following contribution to $G^1$ for every spatial orbital $a$:
$$
	\Delta G^1_{q_{3\alpha} p_{2\beta} p_{1\alpha}, a q_{2\beta} q_1} =
	-\langle p_3 p_4 | q_4 a \rangle
	C_\mathbf{i}C_\mathbf{j}
$$
This process is continued for all the remaining spin-conserving choices for $d, e, c,$ and $f$.


\begin{mcitethebibliography}{37}
\providecommand*\natexlab[1]{#1}
\providecommand*\mciteSetBstSublistMode[1]{}
\providecommand*\mciteSetBstMaxWidthForm[2]{}
\providecommand*\mciteBstWouldAddEndPuncttrue
  {\def\EndOfBibitem{\unskip.}}
\providecommand*\mciteBstWouldAddEndPunctfalse
  {\let\EndOfBibitem\relax}
\providecommand*\mciteSetBstMidEndSepPunct[3]{}
\providecommand*\mciteSetBstSublistLabelBeginEnd[3]{}
\providecommand*\EndOfBibitem{}
\mciteSetBstSublistMode{f}
\mciteSetBstMaxWidthForm{subitem}{(\alph{mcitesubitemcount})}
\mciteSetBstSublistLabelBeginEnd
  {\mcitemaxwidthsubitemform\space}
  {\relax}
  {\relax}

\bibitem[Evangelista(2018)]{doi:10.1063/1.5039496}
Evangelista,~F.~A. \emph{J. Chem. Phys.} \textbf{2018}, \emph{149},
  030901\relax
\mciteBstWouldAddEndPuncttrue
\mciteSetBstMidEndSepPunct{\mcitedefaultmidpunct}
{\mcitedefaultendpunct}{\mcitedefaultseppunct}\relax
\EndOfBibitem
\bibitem[White(1992)]{PhysRevLett.69.2863}
White,~S.~R. \emph{Phys. Rev. Lett.} \textbf{1992}, \emph{69}, 2863--2866\relax
\mciteBstWouldAddEndPuncttrue
\mciteSetBstMidEndSepPunct{\mcitedefaultmidpunct}
{\mcitedefaultendpunct}{\mcitedefaultseppunct}\relax
\EndOfBibitem
\bibitem[Chan and Sharma(2011)Chan, and
  Sharma]{doi:10.1146/annurev-physchem-032210-103338}
Chan,~G. K.-L.; Sharma,~S. \emph{Annu. Rev. Phys. Chem.} \textbf{2011},
  \emph{62}, 465--481\relax
\mciteBstWouldAddEndPuncttrue
\mciteSetBstMidEndSepPunct{\mcitedefaultmidpunct}
{\mcitedefaultendpunct}{\mcitedefaultseppunct}\relax
\EndOfBibitem
\bibitem[Holmes \latin{et~al.}(2017)Holmes, Umrigar, and
  Sharma]{doi:10.1063/1.4998614}
Holmes,~A.~A.; Umrigar,~C.~J.; Sharma,~S. \emph{J. Chem. Phys.} \textbf{2017},
  \emph{147}, 164111\relax
\mciteBstWouldAddEndPuncttrue
\mciteSetBstMidEndSepPunct{\mcitedefaultmidpunct}
{\mcitedefaultendpunct}{\mcitedefaultseppunct}\relax
\EndOfBibitem
\bibitem[Huron \latin{et~al.}(1973)Huron, Malrieu, and
  Rancurel]{doi:10.1063/1.1679199}
Huron,~B.; Malrieu,~J.~P.; Rancurel,~P. \emph{J. Chem. Phys.} \textbf{1973},
  \emph{58}, 5745--5759\relax
\mciteBstWouldAddEndPuncttrue
\mciteSetBstMidEndSepPunct{\mcitedefaultmidpunct}
{\mcitedefaultendpunct}{\mcitedefaultseppunct}\relax
\EndOfBibitem
\bibitem[Evangelisti \latin{et~al.}(1983)Evangelisti, Daudey, and
  Malrieu]{EVANGELISTI198391}
Evangelisti,~S.; Daudey,~J.-P.; Malrieu,~J.-P. \emph{Chem. Phys.}
  \textbf{1983}, \emph{75}, 91 -- 102\relax
\mciteBstWouldAddEndPuncttrue
\mciteSetBstMidEndSepPunct{\mcitedefaultmidpunct}
{\mcitedefaultendpunct}{\mcitedefaultseppunct}\relax
\EndOfBibitem
\bibitem[Wang \latin{et~al.}(2019)Wang, Li, and Lu]{Wang_2019}
Wang,~Z.; Li,~Y.; Lu,~J. \emph{J. Chem. Theory Comput.} \textbf{2019},
  \emph{15}, 3558–3569\relax
\mciteBstWouldAddEndPuncttrue
\mciteSetBstMidEndSepPunct{\mcitedefaultmidpunct}
{\mcitedefaultendpunct}{\mcitedefaultseppunct}\relax
\EndOfBibitem
\bibitem[Booth \latin{et~al.}(2009)Booth, Thom, and
  Alavi]{doi:10.1063/1.3193710}
Booth,~G.~H.; Thom,~A. J.~W.; Alavi,~A. \emph{J. Chem. Phys.} \textbf{2009},
  \emph{131}, 054106\relax
\mciteBstWouldAddEndPuncttrue
\mciteSetBstMidEndSepPunct{\mcitedefaultmidpunct}
{\mcitedefaultendpunct}{\mcitedefaultseppunct}\relax
\EndOfBibitem
\bibitem[Nakatani and Guo(2017)Nakatani, and Guo]{doi:10.1063/1.4976644}
Nakatani,~N.; Guo,~S. \emph{J. Chem. Phys.} \textbf{2017}, \emph{146},
  094102\relax
\mciteBstWouldAddEndPuncttrue
\mciteSetBstMidEndSepPunct{\mcitedefaultmidpunct}
{\mcitedefaultendpunct}{\mcitedefaultseppunct}\relax
\EndOfBibitem
\bibitem[Yanai \latin{et~al.}(2017)Yanai, Saitow, Xiong, Chalupský, Kurashige,
  Guo, and Sharma]{doi:10.1021/acs.jctc.7b00735}
Yanai,~T.; Saitow,~M.; Xiong,~X.-G.; Chalupský,~J.; Kurashige,~Y.; Guo,~S.;
  Sharma,~S. \emph{J. Chem. Theory Comput.} \textbf{2017}, \emph{13},
  4829--4840\relax
\mciteBstWouldAddEndPuncttrue
\mciteSetBstMidEndSepPunct{\mcitedefaultmidpunct}
{\mcitedefaultendpunct}{\mcitedefaultseppunct}\relax
\EndOfBibitem
\bibitem[Wouters \latin{et~al.}(2016)Wouters, Van~Speybroeck, and
  Van~Neck]{Wouters_2016}
Wouters,~S.; Van~Speybroeck,~V.; Van~Neck,~D. \emph{J. Chem. Phys.}
  \textbf{2016}, \emph{145}, 054120\relax
\mciteBstWouldAddEndPuncttrue
\mciteSetBstMidEndSepPunct{\mcitedefaultmidpunct}
{\mcitedefaultendpunct}{\mcitedefaultseppunct}\relax
\EndOfBibitem
\bibitem[Sokolov and Chan(2016)Sokolov, and Chan]{doi:10.1063/1.4941606}
Sokolov,~A.~Y.; Chan,~G. K.-L. \emph{J. Chem. Phys.} \textbf{2016}, \emph{144},
  064102\relax
\mciteBstWouldAddEndPuncttrue
\mciteSetBstMidEndSepPunct{\mcitedefaultmidpunct}
{\mcitedefaultendpunct}{\mcitedefaultseppunct}\relax
\EndOfBibitem
\bibitem[Sharma(2019)]{VMC_PT}
Sharma,~S. \textbf{2019}, arXiv:1803.04341\relax
\mciteBstWouldAddEndPuncttrue
\mciteSetBstMidEndSepPunct{\mcitedefaultmidpunct}
{\mcitedefaultendpunct}{\mcitedefaultseppunct}\relax
\EndOfBibitem
\bibitem[Jeanmairet \latin{et~al.}(2017)Jeanmairet, Sharma, and
  Alavi]{doi:10.1063/1.4974177}
Jeanmairet,~G.; Sharma,~S.; Alavi,~A. \emph{J. Chem. Phys.} \textbf{2017},
  \emph{146}, 044107\relax
\mciteBstWouldAddEndPuncttrue
\mciteSetBstMidEndSepPunct{\mcitedefaultmidpunct}
{\mcitedefaultendpunct}{\mcitedefaultseppunct}\relax
\EndOfBibitem
\bibitem[Andersson \latin{et~al.}(1990)Andersson, Malmqvist, Roos, Sadlej, and
  Wolinski]{doi:10.1021/j100377a012}
Andersson,~K.; Malmqvist,~P.~A.; Roos,~B.~O.; Sadlej,~A.~J.; Wolinski,~K.
  \emph{J. Phys. Chem.} \textbf{1990}, \emph{94}, 5483--5488\relax
\mciteBstWouldAddEndPuncttrue
\mciteSetBstMidEndSepPunct{\mcitedefaultmidpunct}
{\mcitedefaultendpunct}{\mcitedefaultseppunct}\relax
\EndOfBibitem
\bibitem[Andersson \latin{et~al.}(1992)Andersson, Malmqvist, and
  Roos]{doi:10.1063/1.462209}
Andersson,~K.; Malmqvist,~P.~A.; Roos,~B.~O. \emph{J. Chem. Phys.}
  \textbf{1992}, \emph{96}, 1218--1226\relax
\mciteBstWouldAddEndPuncttrue
\mciteSetBstMidEndSepPunct{\mcitedefaultmidpunct}
{\mcitedefaultendpunct}{\mcitedefaultseppunct}\relax
\EndOfBibitem
\bibitem[Finley \latin{et~al.}(1998)Finley, Malmqvist, Roos, and
  Serrano-Andrés]{FINLEY1998299}
Finley,~J.; Malmqvist,~P.-A.; Roos,~B.~O.; Serrano-Andrés,~L. \emph{Chem.
  Phys. Lett.} \textbf{1998}, \emph{288}, 299 -- 306\relax
\mciteBstWouldAddEndPuncttrue
\mciteSetBstMidEndSepPunct{\mcitedefaultmidpunct}
{\mcitedefaultendpunct}{\mcitedefaultseppunct}\relax
\EndOfBibitem
\bibitem[Zobel \latin{et~al.}(2017)Zobel, Nogueira, and González]{C6SC03759C}
Zobel,~J.~P.; Nogueira,~J.~J.; González,~L. \emph{Chem. Sci.} \textbf{2017},
  \emph{8}, 1482--1499\relax
\mciteBstWouldAddEndPuncttrue
\mciteSetBstMidEndSepPunct{\mcitedefaultmidpunct}
{\mcitedefaultendpunct}{\mcitedefaultseppunct}\relax
\EndOfBibitem
\bibitem[Angeli \latin{et~al.}(2002)Angeli, Cimiraglia, and
  Malrieu]{doi:10.1063/1.1515317}
Angeli,~C.; Cimiraglia,~R.; Malrieu,~J.-P. \emph{J. Chem. Phys.} \textbf{2002},
  \emph{117}, 9138--9153\relax
\mciteBstWouldAddEndPuncttrue
\mciteSetBstMidEndSepPunct{\mcitedefaultmidpunct}
{\mcitedefaultendpunct}{\mcitedefaultseppunct}\relax
\EndOfBibitem
\bibitem[Zgid \latin{et~al.}(2009)Zgid, Ghosh, Neuscamman, and
  Chan]{doi:10.1063/1.3132922}
Zgid,~D.; Ghosh,~D.; Neuscamman,~E.; Chan,~G. K.-L. \emph{J. Chem. Phys.}
  \textbf{2009}, \emph{130}, 194107\relax
\mciteBstWouldAddEndPuncttrue
\mciteSetBstMidEndSepPunct{\mcitedefaultmidpunct}
{\mcitedefaultendpunct}{\mcitedefaultseppunct}\relax
\EndOfBibitem
\bibitem[Phung \latin{et~al.}(2016)Phung, Wouters, and
  Pierloot]{doi:10.1021/acs.jctc.6b00714}
Phung,~Q.~M.; Wouters,~S.; Pierloot,~K. \emph{J. Chem. Theory Comput.}
  \textbf{2016}, \emph{12}, 4352--4361\relax
\mciteBstWouldAddEndPuncttrue
\mciteSetBstMidEndSepPunct{\mcitedefaultmidpunct}
{\mcitedefaultendpunct}{\mcitedefaultseppunct}\relax
\EndOfBibitem
\bibitem[Thomas \latin{et~al.}(2015)Thomas, Sun, Alavi, and
  Booth]{doi:10.1021/acs.jctc.5b00917}
Thomas,~R.~E.; Sun,~Q.; Alavi,~A.; Booth,~G.~H. \emph{J. Chem. Theory Comput.}
  \textbf{2015}, \emph{11}, 5316--5325\relax
\mciteBstWouldAddEndPuncttrue
\mciteSetBstMidEndSepPunct{\mcitedefaultmidpunct}
{\mcitedefaultendpunct}{\mcitedefaultseppunct}\relax
\EndOfBibitem
\bibitem[Li~Manni \latin{et~al.}(2016)Li~Manni, Smart, and
  Alavi]{doi:10.1021/acs.jctc.5b01190}
Li~Manni,~G.; Smart,~S.~D.; Alavi,~A. \emph{J. Chem. Theory Comput.}
  \textbf{2016}, \emph{12}, 1245--1258\relax
\mciteBstWouldAddEndPuncttrue
\mciteSetBstMidEndSepPunct{\mcitedefaultmidpunct}
{\mcitedefaultendpunct}{\mcitedefaultseppunct}\relax
\EndOfBibitem
\bibitem[Spencer \latin{et~al.}(2012)Spencer, Blunt, and Foulkes]{Spencer_2012}
Spencer,~J.~S.; Blunt,~N.~S.; Foulkes,~W.~M. \emph{J. Chem. Phys.}
  \textbf{2012}, \emph{136}, 054110\relax
\mciteBstWouldAddEndPuncttrue
\mciteSetBstMidEndSepPunct{\mcitedefaultmidpunct}
{\mcitedefaultendpunct}{\mcitedefaultseppunct}\relax
\EndOfBibitem
\bibitem[Cleland \latin{et~al.}(2010)Cleland, Booth, and
  Alavi]{doi:10.1063/1.3302277}
Cleland,~D.; Booth,~G.~H.; Alavi,~A. \emph{J. Chem. Phys.} \textbf{2010},
  \emph{132}, 041103\relax
\mciteBstWouldAddEndPuncttrue
\mciteSetBstMidEndSepPunct{\mcitedefaultmidpunct}
{\mcitedefaultendpunct}{\mcitedefaultseppunct}\relax
\EndOfBibitem
\bibitem[Blunt(2018)]{doi:10.1063/1.5037923}
Blunt,~N.~S. \emph{J. Chem. Phys.} \textbf{2018}, \emph{148}, 221101\relax
\mciteBstWouldAddEndPuncttrue
\mciteSetBstMidEndSepPunct{\mcitedefaultmidpunct}
{\mcitedefaultendpunct}{\mcitedefaultseppunct}\relax
\EndOfBibitem
\bibitem[Overy \latin{et~al.}(2014)Overy, Booth, Blunt, Shepherd, Cleland, and
  Alavi]{doi:10.1063/1.4904313}
Overy,~C.; Booth,~G.; Blunt,~N.; Shepherd,~J.; Cleland,~D.; Alavi,~A. \emph{J.
  Chem. Phys.} \textbf{2014}, \emph{141}\relax
\mciteBstWouldAddEndPuncttrue
\mciteSetBstMidEndSepPunct{\mcitedefaultmidpunct}
{\mcitedefaultendpunct}{\mcitedefaultseppunct}\relax
\EndOfBibitem
\bibitem[Booth \latin{et~al.}(2012)Booth, Gr{\"u}neis, Kresse, and
  Alavi]{booth/nature}
Booth,~G.~H.; Gr{\"u}neis,~A.; Kresse,~G.; Alavi,~A. \emph{Nature}
  \textbf{2012}, \emph{493}, 365 EP --\relax
\mciteBstWouldAddEndPuncttrue
\mciteSetBstMidEndSepPunct{\mcitedefaultmidpunct}
{\mcitedefaultendpunct}{\mcitedefaultseppunct}\relax
\EndOfBibitem
\bibitem[Blunt \latin{et~al.}(2015)Blunt, Smart, Booth, and Alavi]{Blunt/2015}
Blunt,~N.~S.; Smart,~S.~D.; Booth,~G.~H.; Alavi,~A. \emph{J. Chem. Phys.}
  \textbf{2015}, \emph{143}, 134117\relax
\mciteBstWouldAddEndPuncttrue
\mciteSetBstMidEndSepPunct{\mcitedefaultmidpunct}
{\mcitedefaultendpunct}{\mcitedefaultseppunct}\relax
\EndOfBibitem
\bibitem[Blunt \latin{et~al.}(2017)Blunt, Booth, and
  Alavi]{doi:10.1063/1.4986963}
Blunt,~N.~S.; Booth,~G.~H.; Alavi,~A. \emph{J. Chem. Phys.} \textbf{2017},
  \emph{146}, 244105\relax
\mciteBstWouldAddEndPuncttrue
\mciteSetBstMidEndSepPunct{\mcitedefaultmidpunct}
{\mcitedefaultendpunct}{\mcitedefaultseppunct}\relax
\EndOfBibitem
\bibitem[Blunt \latin{et~al.}(2015)Blunt, Smart, Kersten, Spencer, Booth, and
  Alavi]{doi:10.1063/1.4920975}
Blunt,~N.~S.; Smart,~S.~D.; Kersten,~J. A.~F.; Spencer,~J.~S.; Booth,~G.~H.;
  Alavi,~A. \emph{J. Chem. Phys.} \textbf{2015}, \emph{142}, 184107\relax
\mciteBstWouldAddEndPuncttrue
\mciteSetBstMidEndSepPunct{\mcitedefaultmidpunct}
{\mcitedefaultendpunct}{\mcitedefaultseppunct}\relax
\EndOfBibitem
\bibitem[nec()]{neci}
See https://github.com/ghb24/neci\_stable for NECI github web page.
  \url{https://github.com/ghb24/NECI\_STABLE}\relax
\mciteBstWouldAddEndPuncttrue
\mciteSetBstMidEndSepPunct{\mcitedefaultmidpunct}
{\mcitedefaultendpunct}{\mcitedefaultseppunct}\relax
\EndOfBibitem
\bibitem[Sun \latin{et~al.}(2017)Sun, Berkelbach, Blunt, Booth, Guo, Li, Liu,
  McClain, Sayfutyarova, Sharma, Wouters, and Chan]{pyscf}
Sun,~Q.; Berkelbach,~T.~C.; Blunt,~N.~S.; Booth,~G.~H.; Guo,~S.; Li,~Z.;
  Liu,~J.; McClain,~J.~D.; Sayfutyarova,~E.~R.; Sharma,~S.; Wouters,~S.;
  Chan,~G. K.-L. PySCF: the Python-based simulations of chemistry framework.
  2017; \url{https://onlinelibrary.wiley.com/doi/abs/10.1002/wcms.1340}\relax
\mciteBstWouldAddEndPuncttrue
\mciteSetBstMidEndSepPunct{\mcitedefaultmidpunct}
{\mcitedefaultendpunct}{\mcitedefaultseppunct}\relax
\EndOfBibitem
\bibitem[Shiozaki(2018)]{doi:10.1002/wcms.1331}
Shiozaki,~T. \emph{Wiley Interdiscip. Rev. Comput. Mol. Sci.} \textbf{2018},
  \emph{8}, e1331\relax
\mciteBstWouldAddEndPuncttrue
\mciteSetBstMidEndSepPunct{\mcitedefaultmidpunct}
{\mcitedefaultendpunct}{\mcitedefaultseppunct}\relax
\EndOfBibitem
\bibitem[Blunt \latin{et~al.}(2015)Blunt, Alavi, and
  Booth]{PhysRevLett.115.050603}
Blunt,~N.~S.; Alavi,~A.; Booth,~G.~H. \emph{Phys. Rev. Lett.} \textbf{2015},
  \emph{115}, 050603\relax
\mciteBstWouldAddEndPuncttrue
\mciteSetBstMidEndSepPunct{\mcitedefaultmidpunct}
{\mcitedefaultendpunct}{\mcitedefaultseppunct}\relax
\EndOfBibitem
\bibitem[Blunt \latin{et~al.}(2018)Blunt, Alavi, and Booth]{PhysRevB.98.085118}
Blunt,~N.~S.; Alavi,~A.; Booth,~G.~H. \emph{Phys. Rev. B} \textbf{2018},
  \emph{98}, 085118\relax
\mciteBstWouldAddEndPuncttrue
\mciteSetBstMidEndSepPunct{\mcitedefaultmidpunct}
{\mcitedefaultendpunct}{\mcitedefaultseppunct}\relax
\EndOfBibitem
\bibitem[Rode and Werner(2005)Rode, and Werner]{Werner_cu2o2}
Rode,~M.; Werner,~H.-J. \emph{Theor. Chem. Acc.} \textbf{2005}, \emph{114},
  309--317\relax
\mciteBstWouldAddEndPuncttrue
\mciteSetBstMidEndSepPunct{\mcitedefaultmidpunct}
{\mcitedefaultendpunct}{\mcitedefaultseppunct}\relax
\EndOfBibitem
\end{mcitethebibliography}
\end{document}